\documentclass[journal, twoside, final]{IEEEtran}
\usepackage{cite}
\usepackage{graphicx}
 \usepackage{diagbox} 
\usepackage{amssymb, amsmath, amsthm}
\usepackage{amsfonts}
\usepackage{multirow}
\usepackage{mathrsfs}
\usepackage{color}
\usepackage{url}
\usepackage{flushend}
\usepackage{setspace}
\usepackage{placeins}
\usepackage{booktabs, multicol, multirow}
\usepackage{makecell} 
\usepackage{acronym}
\usepackage{bm}
\usepackage{placeins}
\usepackage{extarrows}
\usepackage{enumitem}
\usepackage{verbatim} 
\usepackage{manfnt}
\allowdisplaybreaks

\usepackage[caption=false]{subfig}

\usepackage{algorithm, algorithmic}

\newtheorem{lemma}{Lemma}
\newtheorem{theorem}{Theorem}
\newtheorem{remark}{Remark}

\newcommand{\RNum}[1]{\uppercase\expandafter{\romannumeral #1\relax}}

\newmuskip\pFqmuskip
\newcommand*\pFq[6][8]{%
  \begingroup 
  \pFqmuskip=#1mu\relax
  \mathcode`\,=\string"8000
  \begingroup\lccode`\~=`\,
  \lowercase{\endgroup\let~}\pFqcomma
  {}_{#2}F_{#3}{\left[\genfrac..{0pt}{}{#4}{#5};#6\right]}%
  \endgroup
}
\newcommand{\pFqcomma}{\mskip\pFqmuskip}

\begin{document}

\title{Ground-to-Air Communications Beyond 5G: Coordinated Multi-Point Transmission Based on Poisson-Delaunay Triangulation}

\author{Yan Li and Minghua Xia, \IEEEmembership{Senior~Member,~IEEE}
\thanks{Manuscript received November 24, 2021; revised May 8, 2022, and July 21, 2022; accepted September 7, 2022. This work was supported in part by the National Natural Science Foundation of China under Grants 62171486 and U2001213, in part by the Natural Science Foundation of Hunan Province under Grant 2022JJ40511, and in part by the Scientific Research Fund of Hunan Provincial Education Department under Grant 21C0180. The associate editor coordinating the review of this paper and approving it for publication was M. Guillaud. {\it (Corresponding author: Minghua Xia.)}}
	\thanks{Yan Li was with the School of Electronics and Information Technology, Sun Yat-sen University, Guangzhou 510006, China; she is now with the School of Computer and Communication Engineering, Changsha University of Science and Technology, Changsha 410004, China (e-mail: liyan228@mail2.sysu.edu.cn).}
	\thanks{Minghua Xia is with the School of Electronics and Information Technology, Sun Yat-sen University, Guangzhou 510006, China, and also with the Southern Marine Science and Engineering Guangdong Laboratory, Zhuhai 519082, China (e-mail: xiamingh@mail.sysu.edu.cn).}
	\thanks{%
	Color versions of one or more of the figures in this article are available online at https://ieeexplore.ieee.org.
	
	Digital Object Identifier XXX}
}

\markboth{IEEE Transactions on Wireless Communications} %
{Li \MakeLowercase{\textit{et al.}}: Ground-to-Air Communications Beyond 5G}

\maketitle

\IEEEpubid{\begin{minipage}{\textwidth} \ \\[12pt] \centering 1536-1276 \copyright\ 2022 IEEE. Personal use is permitted, but republication/redistribution requires IEEE permission. \\
See \url{https://www.ieee.org/publications/rights/index.html} for more information.\end{minipage}}

\IEEEpubidadjcol

\begin{abstract}
\noindent 
This paper designs a novel ground-to-air communication scheme to serve unmanned aerial vehicles (UAVs) through legacy terrestrial base stations (BSs). In particular, a tractable coordinated multi-point (CoMP) transmission based on the geometry of Poisson-Delaunay triangulation is developed, which provides reliable and seamless connectivity for UAVs. An effective dynamic frequency allocation scheme is designed to eliminate inter-cell interference by using the theory of circle packing. For exact performance evaluation, the handoff probability of a typical UAV is characterized, and then the coverage probability with handoffs is attained. Simulation and numerical results corroborate that the proposed scheme outperforms the conventional CoMP scheme with three nearest cooperating BSs in terms of handoff and coverage probabilities. Moreover, as each UAV has a fixed and unique CoMP BS set, it avoids the real-time dynamic BS searching process, thus reducing the feedback overhead.
\end{abstract}


\acrodef{3GPP}{}
\acrodef{CoMP}{}
\acrodef{LTE}{Long-Term Evolution}
\acrodef{UE}{user equipment}


\section{Introduction}
\label{Section:Introduction}
\IEEEPARstart{U}{nmanned} aerial vehicles (UAVs) are widely applied for aerial surveillance, package delivery, and communication relays. Among such applications, UAVs can either act as information providers like aerial base stations (BSs), access points, and cooperating relays, or as aerial user equipments (UEs). According to their distinct roles, the corresponding network paradigms refer to UAV-assisted wireless communications or cellular-connected UAV communications \cite{8470897}. In recent years, cellular-connected UAV communications have been intently studied to fully use the legacy terrestrial cellular systems and provide reliable connections to UAVs. Despite appealing advantages like ubiquitous accessibility, ease of monitoring and management, and cost-effectiveness, cellular-connected UAV communications still face many challenges, such as effective connectivity with terrestrial BSs, interference mitigation, ground-to-air channel modeling, and handoff management \cite{8660516, 8692749,8337920}. In this paper, we will develop a mathematically tractable cellular-connected UAV communication scheme and devise effective handoff management and interference mitigation strategies.

\IEEEpubidadjcol
\subsection{Related Works and Motivation}
Based on system-level simulations or measurement trials, the feasibility of communication connectivity for UAVs by LTE networks was evaluated in \cite{8337920}. Considering 5G technology, the potential use of massive multiple-input multiple-output (MIMO) antennas to support cellular-connected UAVs co-existing with terrestrial UEs was studied in \cite{8869706}. The theory of stochastic geometry was widely exploited to characterize the performance of cellular-connected UAVs. For instance, the downlink coverage probability in a network including static UAVs and terrestrial UEs was analyzed in \cite{8269068}. The downlink coverage probability for UAVs in vertical heterogeneous networks was analyzed in \cite{9013981}, where the coverage probability was disclosed as highly dependent on the UAV’s height. Assuming line-of-sight links, an exact coverage probability of cellular-connected static UAVs was derived in \cite{8422685}. It also reveals that noise and non-line-of-sight links have a negligible effect on the performance of UAVs compared with terrestrial UEs. 

On the other hand, the optimization theory was widely applied to acquire optimal performance of cellular-connected UAVs. For instance, the energy efficiency of the rate-splitting multiple access and non-orthogonal multiple access (NOMA) schemes for UAVs was investigated in \cite{8756699}, and the precoding vectors of both methods were optimized accordingly.  In \cite{8811738}, the weighted sum rate for both UAVs and terrestrial UEs was maximized by jointly optimizing the UAV’s uplink cell associations and the power allocations. Recently, concerning NOMA, the uplink precoding for UAVs was optimized in \cite{8906143}. Furthermore, to effectively mitigate inter-cell interference (ICI) to improve the connectivity of UAVs, coordinated multi-point (CoMP) transmission technology comes into sight. In particular, a new cooperative interference cancellation strategy for UAV uplink transmission was proposed in \cite{8763928}. A closed-form upper bound on the coverage probability of a UAV was derived in \cite{8885505}. Most recently, a CoMP scheme based on binomial-Delaunay tetrahedralization was proposed in \cite{9151343}, where four aerial BSs serve each UAV within the corresponding tetrahedral cells simultaneously. However, these papers mentioned above do not account for the mobility of UAVs for ease of mathematical tractability.

Considering UAVs' mobility, a trajectory scheme was designed in \cite{8531711}, subject to the connectivity constraints, including the minimum received signal-to-noise ratio (SNR), the maximum mobility speed, and the initial and final locations. An iterative sub-channel allocation and speed optimization algorithm was developed to maximize the uplink sum rate in \cite{8624565}. To minimize its mission completion time, the optimal trajectory design for UAVs was studied in \cite{9005434}. Moreover, to reduce the communication latency of UAVs and their interference with terrestrial UEs, an interference-aware path planning scheme was designed in \cite{8654727}, using machine learning. To minimize the total outage duration, a new trajectory design based on deep reinforcement learning was performed in \cite{9406852}. Recently, to rapidly model and evaluate the performance of mobile UAVs,  two classical spatial mixed mobile models for UAVs with or without pause time were proposed in \cite{8671460} and \cite{8998329}, respectively, inspired by the classical plane mobile model \cite{8673556}. Moreover, using the theory of stochastic geometry,  the coverage probability for terrestrial UEs and UAVs was analytically derived in \cite{8671460} and \cite{8998329}, respectively. In this paper, to jointly consider the tractability of the network model and the mobility of UAVs, the classical model, widely known as the modified random waypoint model, will be adopted.

Once the mobility of UAVs is considered, their handoff performance is imperative for the designed network. Regarding two-dimensional (2D) planar scenarios, classical handoff algorithms are usually based on the maximum received power of users \cite{1105931} or on the user's dwell-time threshold \cite{634792}. There are also speed-based handoff algorithms that can estimate the user's speed and direction to determine whether a handoff occurs or not \cite{587613}. Other handoff criteria include signal power strength and mobile user speed, see, e.g., \cite{556473}. For ease of mathematical tractability for handoff performance evaluation, some approximation methods were introduced to simplify optimization problems to get optimal solutions in closed form. For example, given the sojourn time distribution, an effective mobile model was proposed, and the user transition probability from different cells was derived in \cite{622908}. Meanwhile, given the location of terrestrial BSs, the movement trajectory of UEs, and the channel state information, an upper bound on the system handoff performance was studied in \cite{7725961}. 

Regarding three-dimensional (3D) spatial scenarios, there is little literature focusing on UAVs' handoff performance. Using the Poisson point process (PPP) to describe the distribution of terrestrial BSs, each projected UAV is associated with its nearest BS, and the resulting polygonal boundaries around BSs form a Poisson-Voronoi tessellation \cite{Haenggi12}. Based on the Poisson-Voronoi tessellation, the mobile handoff probability for UAVs was studied in \cite{8673556}. Using CoMP, the cooperative handoff performance for the projected UAV inside conventional hexagonal grids was analytically investigated in \cite{8998329}. However, the assumption of such regular cells does not coincide with the practical deployment of BSs. Insofar as a mathematically tractable cellular-connected UAV communication architecture integrating CoMP and handoffs is still in its infancy.

The motivation of this paper comes from the works \cite{Xia2018Un, 8976426}, where the Poisson-Delaunay triangulation was exploited to model the 2D cellular networks for terrestrial UEs. A similar idea is employed in this paper to provide service for aerial UAVs, which benefits by maximizing the use of existing cellular network infrastructure and providing reliable communication connectivity for UAVs, in addition to terrestrial UEs. In particular, in this paper, the PPP is applied to model the locations of terrestrial BSs, and the Delaunay tessellation is exploited for cellular network modeling, yielding a network model called Poisson-Delaunay triangulation. Moreover, a frequency-allocation scheme is designed to eliminate ICI by using the theory of circle packing, and its performance is evaluated by the high-dimensional probability theory. Compared with the conventional CoMP strategy with three nearest cooperating BSs \cite{NigamTCOM14s}, simulation and numerical results corroborate that the UAV handoff probability in the proposed scheme reduces to about two-thirds of that in \cite{NigamTCOM14s}.  The main reason behind the reduction is that the nearest distance criterion adopted by the former is more sensitive to UAV mobility.

\subsection{Contributions}
In this paper, we first develop a mathematically tractable cellular-connected UAV communication architecture based on CoMP transmission. Then, the handoff and coverage probabilities for mobile UAVs are analytically derived. Afterward, a practical frequency allocation scheme is designed to eliminate ICI effectively. In summary, the main contributions of this paper are threefold:
\begin{enumerate}[label = {\arabic*)}]
	\item Network modeling: A cellular-connected UAV communication architecture based on Poisson-Delaunay triangulation is proposed, where CoMP transmission for UAVs is applied to enhance communication connectivity. In particular, each UAV is concurrently served by three terrestrial BSs forming a triangular cell. 
	
	\item Handoff management and coverage probability analysis: Taking advantage of equivalent circle approximation, a simple but effective handoff management mechanism is first developed. Then, the handoff probability and coverage probability with handoffs for mobile UAVs are derived. Numerical and simulation results show that the handoff probability of the proposed scheme is much lower than that of the conventional CoMP scheme with three nearest cooperating BSs.
		
	\item Frequency planning strategy: To effectively eliminate ICI, a dynamic frequency allocation algorithm is designed by means of circle packing, and its performance is evaluated using the theory of high-dimensional probability.
\end{enumerate}

\subsection{Paper Organization}
To detail the contributions mentioned above, the remainder of this paper is organized as follows. Section~\ref{Section_SystemModel} describes the system model. Section~\ref{Section_PerformanceAnalysis} analyzes the performance of a typical UAV in terms of handoff probability and coverage probability with handoffs. Section~\ref{Section_FrequencyAllocation} develops a dynamic frequency allocation strategy. Simulation results are presented and discussed in Section~\ref{Section_Simulation} and, finally, Section~\ref{Section_Conclusion} concludes the paper.

{\it Notation}: The operator ${\rm round}(\cdot)$ gives the nearest integer of a real number, and $\lfloor \cdot \rfloor$ returns the maximum integer not larger than a real number. The operator $\mathbb{E}(\cdot)$ means mathematical expectation. The notation $\|\bm{x}\|$ and $\bm{x}^H$ denote the $\ell_2$-norm and Hermitian transpose of vector $\bm{x}$, respectively. The symbol $ {n \choose m} \triangleq \frac{n!}{m! \, (n-m)!}$ indicates binomial coefficient, with $n!$ being the factorial of an integer $n$. The Gamma function is defined as $\Gamma(a) \triangleq \int_0^\infty{x^{a-1} \exp(-x)}\,{\rm d}x$ for $a>0$. The lower incomplete Gamma function is defined as $\gamma(a, x) \triangleq \int_{0}^{x}t^{a-1}\exp(-t) {\rm d}t$ and $\Gamma(v, \theta)$ denotes a Gamma distribution with shape parameter $v$ and scale factor $\theta$. The generalized hypergeometric function ${\displaystyle \, {}_{p}F_{q}(a_{1},\cdots ,a_{p}; b_{1}, \cdots, b_{q}; x) \triangleq \sum_{n=0}^{\infty }{\frac {(a_{1})_{n} \cdots (a_{p})_{n}}{(b_{1})_{n} \cdots (b_{q})_{n}}} \, {\frac {x^{n}}{n!}}}$, with $(a)_n = a(a+1)\cdots(a+n-1)$ if $n > 1$ and $(a)_n = 1$ if $n = 0$. Notice that these special functions can be readily computed using built-in functions in regular numerical software, such as Matlab and Mathematica. 

\begin{figure}[!t] 
	\centering    
	\includegraphics[width=3.0in, clip, keepaspectratio]{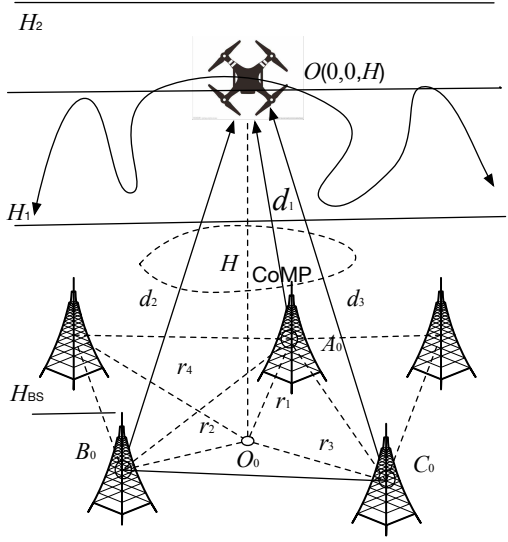}	
	\caption{An illustration of the proposed network based on Poisson-Delaunay triangulation, where a typical UAV flying at the height $H$ ($H_1 \le H \le H_2$) is served by the BSs in the CoMP set $\{A_0, B_0, C_0\}$ with distances $d_i, i \in \{1,2,3\}$. } 
	\label{Fig_1}      
\end{figure}

\section{System Model}
\label{Section_SystemModel}
As shown in Fig.~\ref{Fig_1}, we consider a cellular-connected UAV network where terrestrial BSs are assumed to be distributed as a homogeneous PPP, denoted $\Phi$, with intensity $\lambda$. Each BS is equipped with $M$ antennas and deployed at the same height $H_{{\rm BS}}$. Each UAV has a single antenna and is flying at a height within $[H_1, H_2]$, where $H_2 > H_1 \geq H_{{\rm BS}}$. Without loss of generality, a typical UAV, which is independent of the BS distribution, is assumed to be located at $O(0, 0, H)$ and the projection of $O$ onto the plane is exactly at the origin $O_0(0, 0, 0)$. To enhance the connectivity of UAVs, a typical UAV is served by the three BSs in the CoMP cooperating set, $\Phi_0 = \{A_0, B_0, C_0\}$. The distances of a BS within $\Phi_0$ from $O_0$ and $O$ are denoted by $r_i$ and $d_i = \sqrt{r_i^2+H^2}$, $\forall i \in \{1, 2, 3\}$, respectively.  Next, we elaborate on how to determine the CoMP set of a UAV, followed by mobility and channel models, as well as performance metrics.

\begin{figure}[!t] 
			\centering      
			\includegraphics[width=3.5in, clip, keepaspectratio]{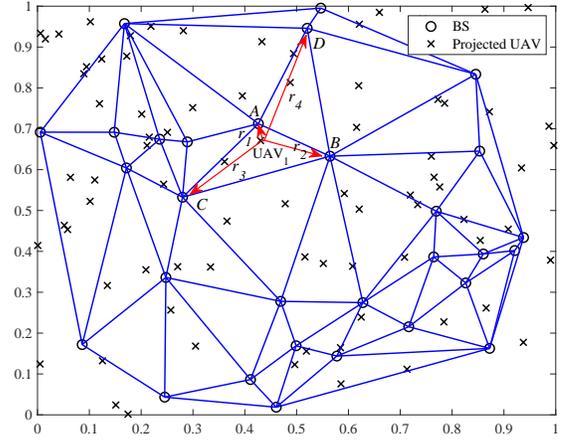}  
	\caption{Poisson-Delaunay triangulation network where the CoMP sets are defined by Delaunay triangles (solid blue boundaries), where BSs (black circles) are distributed as a PPP with a normalized coverage area of one squared kilometer, and the projected UAVs (cross marks) are uniformly distributed on a 2D plane.} 
	\label{Fig-2}
\end{figure}

\subsection{The CoMP Set of a UAV}
\label{The CoMP Set of a UAV}
As a companion to the 3D scenario shown in Fig.~\ref{Fig_1}, Fig.~\ref{Fig-2} shows the BSs and projected UAVs on a 2D plane, denoted by `$\circ$' and `$\times$' marks, respectively. If each projected UAV is associated with its nearest BS, the resulting polygonal boundaries form a Poisson-Voronoi tessellation. The dual Poisson-Delaunay triangulation is illustrated in Fig.~\ref{Fig-2} as the triangles with solid blue boundaries. For more details on the relationship between Poisson-Voronoi tessellation and its dual Poisson-Delaunay triangulation, please refer to \cite{8976426} and the references therein.

Given the geometric locations of BSs, the Poisson-Delaunay triangulation in Fig.~\ref{Fig-2} is uniquely determined and can be efficiently constructed using, e.g., the radial sweep or divide-and-conquer algorithm \cite[ch. 4]{Hjelle06}. Then, for each projected UAV, the CoMP cooperation set can be readily determined. As shown in Fig.~\ref{Fig-2}, for the projected ${\rm UAV}_1$, it firstly chooses the nearest BS $A$ and the second nearest BS $B$ as its two associated BSs, thus determining an edge of a cooperative triangular cell. There must be two adjacent triangles that share the same edge and form a quadrilateral $\{A, B, C, D\}$. Among the four BSs at the vertices of the quadrilateral, ${\rm UAV}_1$ chooses the two nearest BSs and a third BS between the remaining two opposite BSs that is closer to ${\rm UAV}_1$ to form the CoMP cooperating set $\{A, B, C\}$. For a typical UAV, the BSs in $\Phi_0$ are called the serving BSs, while all the other BSs are regarded as interfering ones.

\subsection{Mobility Model}
\begin{figure}[!t]
	\centering
	\includegraphics [width=3.5in, clip, keepaspectratio]{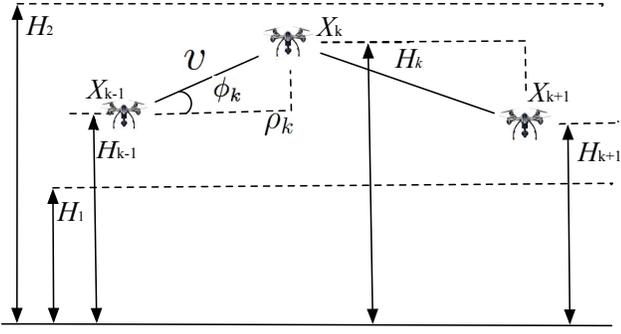}
	\caption{A sample trace of a 3D modified random waypoint mobility model.}
	\label{Fig-3}
\end{figure}

We employ the 3D mobility model widely used in the literature, e.g., in \cite{8998329}, which is directly extended from the modified random waypoint model on a 2D plane \cite{6477064}. As shown in Fig.~\ref{Fig-3}, the movement trace of a UAV can be described by an infinite sequence of tuples: $\{(X_{k-1}, X_{k}, v)\}$, $\forall X_k \in \mathbb{R}^3$, where $k$ is the movement epoch and $X_k$ is the 3D displacement of the UAV at epoch~$k$. During the $k^{\mathrm{th}}$ movement epoch, the UAV is initially from the starting waypoint $X_{k-1}$ at height $H_{k-1}$, then it moves at a constant speed ${v}$ towards the target waypoint $X_{k}$ at height $H_{k}$ with horizontal transition distance $\rho_{k}$ and angle $\phi_{k}$ between the movement direction and the horizontal direction. In general, the horizontal transition length $\rho_{k}$ is characterized by the Rayleigh distribution $f_{\rho_k} (\mu)$ with mobility parameter $\mu$. The heights $H_{k-1}$ and $H_{k}$ are chosen uniformly from the finite height interval $[H_{1}, H_{2}]$. The selection of waypoints is assumed to be independent and identical for each movement epoch, and there is no pause time at these waypoints \cite{article}. Once the UAV reaches $X_{k}$, it repeats the same procedure to find the next target waypoint $X_{k+1}$ at height $H_{k+1}$, and so on. Let $H$ denote the height of the UAV at a steady state; then its PDF is given by \cite{8998329}
    \begin{align}\label{Eq_Height_PDF}
  f_{H}(x) &=\frac{6}{(H_{2}-H_{1})^3} \left(H_{1} x +H_{2} x- H_{1}H_{2} -x^2\right), \nonumber\\
  &\qquad H_{1}< x<H_{2}.
  \end{align}
  If $H_{1} = 0$, then \eqref{Eq_Height_PDF} reduces to
  \begin{equation}
   f_{H}(x) = \frac{6}{H_2^3}\left(H_2 x-x^2\right), \quad 0<x<H_2,
  \end{equation}
  which is in accordance with \cite[Eq. (3)]{8671460}.

 \subsection{Channel Model}
Suppose that all BSs transmit signals with the same power level $P$, then the signal power received at a typical UAV from the three BSs in the CoMP set, $\Phi_0 = \{A_0, B_0, C_0\}$, can be computed as $S = \left|\sum_{i \in \Phi_0}P^{\frac{1}{2}} d^{-\frac{\alpha}{2}}_i \bm{h}^{H}_i\bm{\omega}_i \right|^2$, where $d_i$ indicates the Euclidean distance from the $i^{{\rm th}}$ BS to a typical UAV (cf. Fig.~\ref{Fig_1}); $\bm{h}_i \in \mathbb{C}^{M \times 1}$ represents the complex-valued channel vector from the $i^{\mathrm {th}}$ serving BS to a typical UAV; $\bm{\omega}_i \in \mathbb{C}^{M \times 1}$ is the precoder used at the $i^{{\rm th}}$ BS, and $\alpha > 2$ refers to the path-loss exponent. Likewise, the interference power can be computed as $I = \sum_{j \in \Phi \setminus \Phi_0}P d^{-\alpha}_{j,0}|\bm{h}^{H}_{j, 0}\bm{\omega}_j|^2$, where the difference set $\Phi \setminus \Phi_0$ represents the point process of interfering BSs; $d_{j, 0}$ and $\bm{h}_{j, 0}$ stand for the distance and channel vector from the $j^{{\rm th}}$ interfering BS to a typical UAV, respectively. 

For general purposes, Nakagami-$m$ fading is assumed throughout the paper to capture a large class of fading environments. Also, full downlink channel state information (CSI) is assumed to be available at BSs. Accordingly, the channel-inverse precoder $\bm{\omega}_i$ used by the $i^{{\rm th}}$ BS is given by $\bm{\omega}_i = {\bm{h}_i}/{\|\bm{h}_i\|}$. Since $\bm{h}_i \in \mathbb{C}^{M \times 1}$ is a random vector with each element $h_k$, $k\in \{1,2\cdots, M\}$, being complex-valued Gaussian variable $h_k \sim \mathcal{CN}(\sqrt{K}, 1)$\footnote{The parameter $K$ refers to Ricean factor and $\sqrt{K}$ is exactly the mean of $h_k$. The relation between the Ricean factor $K$ and the Nakagami shape factor $m$ is specified in \cite[Eq. (2.66)]{Gordon2017Principles}.}, $|h_k|^2$ follows the Gamma distribution with probability density function (PDF) $f(x) = \left({m}/{\Omega}\right)^m {x^{m-1}}\exp\left(-{m}/{\Omega}x\right)/{\Gamma(m)}$ \cite[Eq. (2.67)]{Gordon2017Principles} where $m \approx (K+1)^2/(2K+1)$ and $\Omega = K+1$. Thus, the channel gain $g_i \triangleq |\bm{h}^{H}_i\bm{\omega}_i|^2, i \in \{1,2,3\}$, approximately follows the Gamma distribution $\Gamma(m_1, \Omega_1)$ with $m_1 = mM$ and $\Omega_1 = \Omega/m$, which is the sum of $M$ independent and identically distributed Gamma random variables. As for the interference links, the channel gain $g_j \triangleq |\bm{h}^{H}_{j,0} \bm{\omega}_j|^2$, $j\in \Phi \setminus \Phi_0$, also follows Gamma distribution $\Gamma(m_2, \Omega_2)$, where $m_2$ and $\Omega_2$ can be determined by the second-order moment matching method \cite[Lemma 7]{5953530}. 

\subsection{Performance Metrics}
According to the discussion in the preceding subsection, the received signal-to-interference ratio (SIR) at  a typical UAV can be computed as
\begin{equation}\label{Eq_RxSINR}
\eta \triangleq \frac{S}{I} = \frac{\bigg|\sum\limits_{i \in \Phi_0} d^{-\frac{\alpha}{2}}_i \bm{h}^{H}_i\bm{\omega}_i\bigg|^2}{\sum\limits_{j \in \Phi \setminus \Phi_0} d^{-\alpha}_{j,0}g_j}.
\end{equation}
Then, similar to \cite{7006787, 7866856}, given a SIR threshold $\gamma$, the coverage probability with handoffs is defined as
\begin{align}
\mathcal{P} &\triangleq \Pr \left\{\eta > \gamma, {\rm \overline{Hoff}} \right\} + (1-\beta)\Pr\left\{ \eta > \gamma, {\rm Hoff} \right\} \label{Eq_Coverage_Handoff_form1} \\
&= (1-\beta)\Pr \left\{\eta > \gamma \right\} + \beta \Pr \left\{\eta > \gamma, {\rm \overline{Hoff}} \right\} \nonumber\\
&=\left[(1-\beta)+ \beta \left(1-\mathbb{P}_{\rm {H}} \right) \right]	\mathbb{P}_{\rm{C}}, \label{Eq_Coverage_Handoff_form2}
\end{align}
where in \eqref{Eq_Coverage_Handoff_form1} ${\rm {Hoff}}$ and ${\rm \overline{Hoff}}$ denote the events that a handoff occurs and no handoff occurs, respectively; the factor $\beta \in [0, 1]$ is the probability of connection failure due to handoffs, which reflects the system sensitivity to handoffs. In particular, if $\beta= 0$, then the coverage of a UAV is not sensitive to handoffs and becomes equal to the traditional coverage probability; if $\beta = 1$, however, every handoff leads to an outage. Clearly, the first term on the right-hand-side of \eqref{Eq_Coverage_Handoff_form1} refers to the probability of the joint events that a UAV is in coverage and no handoff occurs, while the second term denotes the probability of the joint events that a UAV is in coverage and a handoff occurs, penalized by the cost of handoff.  Finally, in \eqref{Eq_Coverage_Handoff_form2} $\mathbb{P}_{\rm {H}} \triangleq \Pr\{\rm Hoff \}$ refers to the handoff probability while $\mathbb{P}_{\rm C} \triangleq \Pr \left\{\eta > \gamma \right\}$ denotes the coverage probability without handoff.

\section{Coverage Probability with Handoffs}
\label{Section_PerformanceAnalysis}
In this section, an equivalent Voronoi approximation method is first developed to characterize the handoff area of a typical UAV. Then, an integral expression on the handoff probability is derived. Finally, the coverage probability with handoffs is explicitly formulated.

We begin to define a handoff event in our handoff management scheme. As previously shown in Fig.~\ref{Fig-2}, whether a typical UAV makes a handoff or not depends on whether its associated CoMP set is changed or not. In essence, this corresponds to the change of the average received power for a typical UAV, where the average is taken with respect to the channel fading. Let $\Phi_j = \{A_j, B_j, C_j\}$, $j = 0,1,\cdots$, denotes the $j^{\rm th}$ CoMP set, then $\Phi= \{\Phi_j\}|_{j=0}^{\infty}$. It is clear that, if $E_j = \left|\sum_{i \in \Phi_j} P^{\frac{1}{2}} d^{-\frac{\alpha}{2}}_i \right|^2< E_k = \left|\sum_{i \in \Phi_k} P^{\frac{1}{2}} d^{-\frac{\alpha}{2}}_{i} \right|^2$, then a handoff from the $j^{\rm th}$ CoMP set to its adjacent $k^{\rm th}$ one occurs. 

\subsection{The Strategy of Handoff Management}
\label{Subsection-Handoff}
Since a handoff involve possible changes of one, two, or three CoMP BSs within a Delaunay triangle (cf. Fig.~\ref{Fig-2}), it is hard to accurately characterize the coverage area of a {\it virtual cell}, namely, the geographic region where a UAV associates with the same CoMP set of BSs before any handoff occurs. As a result, this hinders us from deriving the handoff probability, which is the probability that a UAV moves across adjacent virtual cells during one movement period. To proceed with handoff management, we propose to approximate virtual cells using the dual Voronoi cells. 

\begin{figure}[t!] 
	\centering    
	\subfloat[Illustration of four virtual cells.] 
	{
		\begin{minipage}[t]{0.5\textwidth}
			\centering         
			\includegraphics[width=1.0\textwidth]{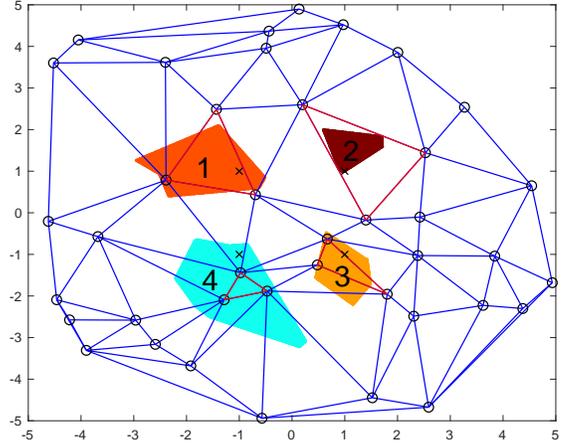}  
			\vspace{-20pt}
			\label{Fig-4a}
		\end{minipage}
	}

	\subfloat[The Voronoi approximation method.] 
	{
		\begin{minipage}[t]{0.5\textwidth}
			\centering      
			\includegraphics[width= 1.0\textwidth]{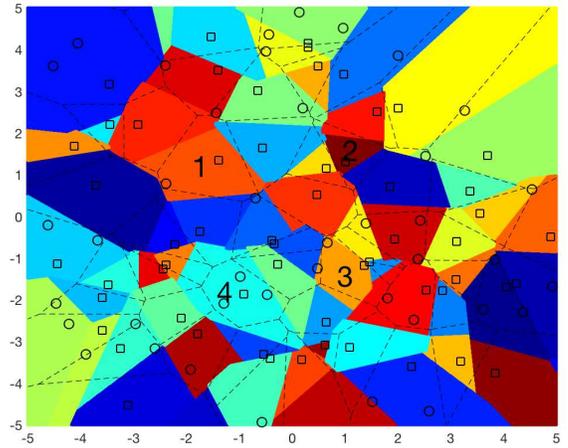}  
			\vspace{-20pt}
			\label{Fig-4b} 
		\end{minipage}
	}
	\caption{An illustration of virtual cells and the corresponding Voronoi approximation. The shaded areas with different colors in (b) represent the original virtual cells, while the polygons with black dashed boundaries denote the newly formed Voronoi tessellation w.r.t. the black squares located at the centers of the circumcenters of triangular cells in (a).} 
\end{figure}

Now, we explain our handoff management strategy from a geometry perspective. For illustration purposes, Fig.~\ref{Fig-4a} plots four {\it virtual cells} with distinct colors, each denoting the geographic region where a UAV associates with the same CoMP set of BSs before any handoff occurs. For a UAV moving within each virtual cell, it is always served by the three BSs at the vertices of the corresponding red triangular cell, and no handoff occurs. Otherwise, a handoff between adjacent cells occurs. On the other hand, Fig.~\ref{Fig-4b} shows the Voronoi tessellation (i.e., the dotted polygons) formed by the black square marks that locate at the centers of the circumcircles of the triangular cells in Fig.~\ref{Fig-4a}. It is observed that the dotted polygons in Fig.~\ref{Fig-4b} can reasonably approximate the corresponding shaded regions with different colors, which represent the virtual cells of our handoff management. Moreover, the area distribution of the dotted polygons in Fig.~\ref{Fig-4b} can be easily computed by that of the Delaunay triangular cells in Fig.~\ref{Fig-4a} by using the theory of stochastic geometry. Consequently, the dotted polygons formed by the nearest-neighbor criterion with respect to the black squares shown in Fig.~\ref{Fig-4b} are employed to analyze the handoff probability, as detailed below.

	\begin{remark}[On the approximation of virtual cells by the dual Voronoi cells]
	In theory, from a geometrical viewpoint, it is clear that the virtual cells in Fig.~\ref{Fig-4a} and the approximate Voronoi cells in Fig.~\ref{Fig-4b} are formed by the same PPP, depicted by the black squares in Fig.~\ref{Fig-4b}. As the average area of polygonal Voronoi cells depends only upon the density of its underlying PPP \cite[Table 5.5.1]{b:OkabeBoots2000}, the average area of virtual cells in Fig.~\ref{Fig-4a} is equal to that of the dual Voronoi cells in Fig.~\ref{Fig-4b}. In practice, however, the average energy received by a typical UAV in virtual cells is not exactly but almost equal to that in the dual Voronoi cells. Let's take $\rm{UAV}_1$ in Fig.~\ref{Fig-2} for instance. By the criterion to determine the CoMP set described in Section II-A, the BSs $A$ and $B$ in the CoMP set of $\rm{UAV}_1$ must be the two nearest ones, but the BS $C$ is not necessarily the third nearest. In contrast, as per the nearest-neighbor criterion, the distance between a UAV and its serving BS in the dual Voronoi cells is always the closest. As a result, this inaccuracy slightly overestimates the handoff probability, as later shown in Fig.~\ref{Fig-7}.
	\end{remark}

\subsection{Handoff Probability}
\label{Sec_HandoffProbability}
\begin{figure}[!t]
	\centering
	\includegraphics [width=2.8in, clip, keepaspectratio]{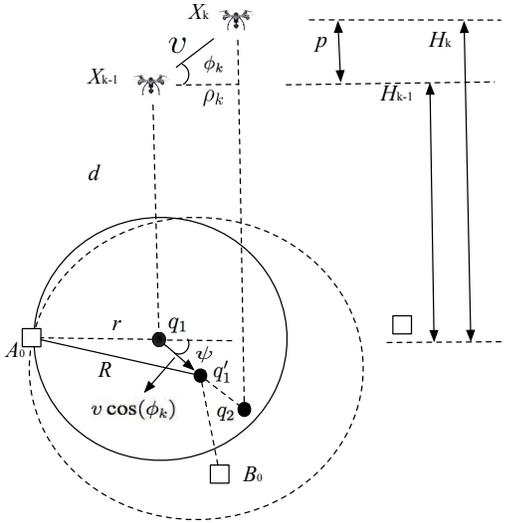}
	\caption{Delaunay triangular handoff scenario, the UAV starts from the projected location $q_{1}$, at connection distance $r$ from the equivalent serving BS $A_0$, moving a distance ${v}\cos(\phi_k)$ in the unit of time at angle $\psi$ with the direction of the connection. The handoff occurs if another BS is closer than $R$ to the projected UAV at the new location $q_1^{\prime}$.}
	\label{Fig-5}
\end{figure}

As illustrated in Fig.~\ref{Fig-5}, define $q_{1}$ and $q_2$ in $\mathbb{R}^2$ as the horizontal projections of two consecutive waypoints $X_{k-1}(\rho_{k-1}, \psi_{k-1}, H_{k-1})$ and $X_{k}(\rho_{k}, \psi_{k}, H_{k})$, respectively. The UAV starting from the projected location $q_{1}$ that is at a distance $r$ from the serving BS $A_0$, moves at a speed $v$ to $q_{1}^{\prime}$ at a distance ${v}\cos(\phi_k)$ in a unit time. Suppose that the UAV does not change its direction during any movement epoch, then $q_{1}^{\prime}$ lies in the segment from $q_{1}$ to $q_2$. By using the law of cosines, $q_{1}^{\prime}$ is at distance $R=\left(r^2+({v}\cos(\phi_k))^2+2r {v}\cos(\phi_k) \cos(\psi) \right)^{1/2}$ from the BS $A_0$, where $\psi$ denotes the angle between the horizontal direction and the direction from $q_1$ to $q_1^{\prime}$ (cf. Fig.~\ref{Fig-5}). As a result, a handoff occurs if there is another BS (e.g., $B_0$ in Fig.~\ref{Fig-5}) closer than $R$ to the projected UAV at the new location $q_1^{\prime}$. Since a typical UAV moves a distance ${v}\cos(\phi_k)$ after a unit time, the conditional handoff probability can be formulated in the following lemma.

\begin{lemma} \label{Lemma_1}
	Given $\{\rho_{k}, H_{k-1}, H_k\}$, the conditional handoff probability can be calculated as 
	\begin{align} \label{Eq_CondotionalHandoffPr_2}
		& {\Pr}_{{\rm H}|\{\rho_{k},H_{k-1}, H_k \}}\left\{E_0 < E_j \right\} \nonumber \\
		& \quad = \left\{\begin{array}{rl}
			1-\frac{1}{\pi}l_1, &  \text{if } {v}\cos(\phi_k) \leq r; \\ 
			1-\frac{1}{\pi}(l_2+l_3+l_4), & \text{otherwise},
		\end{array} \right. 
	\end{align}
where 
$l_1 \triangleq \int_{0}^{\infty}\int_{0}^{\pi} f(u_k, r, \psi) {\rm d}\psi \, {\rm d}r$, 
$l_2 \triangleq \int_{0}^{\infty}\int_{0}^{{\pi}/{2}}f(u_k, r,\psi) {\rm d}\psi \, {\rm d}r$, 
$l_3 \triangleq \int_{0}^{\zeta}\int_{{\pi}/{2}}^{\pi} f(\pi-u_k, r,\psi) {\rm d}\psi \, {\rm d}r$, and
$l_4 \triangleq \int_{\zeta}^{\infty}\int_{{\pi}/{2}}^{\pi} f(u_k, r,\psi)  {\rm d}\psi \, {\rm d}r$,
with $u_k \triangleq \pi-\psi+\arcsin\left({{v}\cos(\phi_k)\sin(\psi)}/{R} \right)$, $\zeta \triangleq {v}\cos(\phi_k)\cos(\pi-\psi)$, 
and $f(u_k, r, \psi) \triangleq \exp\left(-2\lambda \left(R^2 u_k -r^2(\pi-\psi)+r{v}\cos(\phi_k)\sin(\psi)  \right) \right) f_r(r)$ in which $f_r(r)$ denotes the PDF of the nearest distance between the projected UAV and the equivalent serving BSs denoted by black squares in Fig.~\ref{Fig-5}, given by \cite{AndrewsTCOM1111}
\begin{equation} \label{Eq_PDF_d}
	f_r(r) = 4 \lambda \pi r \exp\left(-2\lambda \pi r^2\right).
	\end{equation} 
\end{lemma}
\begin{proof}
	See Appendix~\ref{Proof_Lemma_1}.
\end{proof}

In light of Lemma~\ref{Lemma_1}, by taking the average with respect to $\rho_{k}$, $H_{k-1}$, and $H_k$, the handoff probability can be computed as 
\begin{equation}\label{Eq_HandoffPr}
\mathbb{P}_{\rm H} = \hspace{-5pt}\int\limits_{0}^{\infty}\int\limits_{H_{1}-H_{2}}^{H_{2}-H_{1}} \hspace{-8pt} \mathbb{P}_{\rm H|\{\rho_{k},H_{k-1}, H_k \}}\left\{E_0 < E_j \right\} f_p(p) f_{\rho_{k}}(\rho_{k})  \, {\rm d}p \, {\rm d}\rho_{k},
\end{equation}
where $p \triangleq H_{k} - H_{k-1}$ denotes the height difference of a UAV between two consecutive waypoints $X_{k-1}(\rho_{k-1}, \psi_{k-1}, H_{k-1})$ and $X_{k}(\rho_{k}, \psi_{k}, H_{k})$, $f_p(p) \triangleq {(H_{2}-H_{1})-|p|}/{(H_{2}-H_{1})^2}$ with $H_{1}-H_{2}<p<H_{2}-H_{1}$, $f_{\rho_k} (\rho_k) \triangleq 2\pi \mu \rho_k \exp(-\pi \mu \rho_k^2)$. Note that $p>0$ corresponds to the rising process of a UAV, whereas $p<0$ to the falling process.

\subsection{Coverage Probability with Handoffs}
\label{Section_CoverageProbability_Handoff}
According to the way that each UAV chooses its cooperating set (cf. Fig.~\ref{Fig-2}), the distances between each UAV and its three serving BSs can be approximated to the first three nearest ones $\{d_1, d_2, d_3\}$. Then, the SIR given by \eqref{Eq_RxSINR} reduces to
\begin{equation} \label{Eq_RxSINR-CaseI} 
	\eta \approx \frac{\left| \sum\limits_{i=1}^{3}d^{-\frac{\alpha}{2}}_i \|{h}_i\| \right|^2}{\sum\limits_{j =4}^{\infty} d^{-\alpha}_{j, \, 0} g_j }.
\end{equation}

By using the theory of Palm distribution, the joint distribution of the horizontal distance parameters $r_1$, $r_2$, and $r_3$ is obtained in \cite{Moltchanov2012Distance}, expressed as
\begin{equation} \label{Eq_JointPDF_TypeI}
	f_{r_{1}, r_{2}, r_{3}}(x, y, z) = (2\lambda \pi)^3 x y z \exp\left(-\lambda \pi z^2\right).
\end{equation}

With the PDF of $H$ shown in \eqref{Eq_Height_PDF} and the joint PDF of $\{r_1, r_2, r_3\}$ given by \eqref{Eq_JointPDF_TypeI}, the coverage probability of a typical UAV can be formalized as follows.
\begin{lemma} \label{Lemma_2}
	With a prescribed outage threshold $\gamma$, the coverage probability of a typical UAV is upper bounded by
	\begin{equation}\label{Eq_Theorem_1}
		\mathbb{P}_{\rm C_1} \leq \int\limits_{H_{1}}^{H_{2}}\int\limits_{0<\bm{r} <\infty} \left \| \exp(\bm{Q}(d_i))  \right \|_1  f_{H}(x) f_{r_{1}, r_{2}, r_{3}}(\bm{r}) \, {\rm d}\bm{r} \, {\rm d}x , 
	\end{equation}
	where $\{\bm{r}: 0 < r_{1} \leq r_{2} \leq r_{3} < \infty\}$ and $\bm{Q}(d_i)$ is a $\varpi \times \varpi$ lower triangular Toeplitz matrix, expressed as
	\begin{equation}\label{Eq_ToplitzQMatrix}
		\bm{Q} = \left[ \begin{matrix}
			q_0\\q_1&q_0\\q_2&q_1&q_0 \\ \vdots & \vdots& \vdots &\ddots \\ q_{\varpi-1}&\cdots &q_2&q_1&q_0
			\end{matrix}\right],
	\end{equation} 
	with $\varpi = {\rm round}(3m_1)$ and the entry $q_k$, $k = 0, 1, \cdots, \varpi-1$, given by
	\begin{align}\label{Eq_q_n}
		q_k &= \lambda \pi {d_3}^{2} \delta(k) -\lambda \pi {d_3}^{2-k\alpha} \left({\Omega_2\gamma}\right)^k\left({\Omega_1 \sum\limits_{i=1}^3 {d_i}^{-\alpha}}\right)^{-k} \nonumber \\
			& \quad \times \frac{2\Gamma(k+m_2)}{k!(2-k\alpha)\Gamma(m_2)} \ \pFq{2}{1}{k+m_2, k-\frac{2}{\alpha}}{k+1-\frac{2}{\alpha}}{-\frac{\Omega_2  \gamma{d_3}^{-\alpha}}{\Omega_1 \sum\limits_{i=1}^3 {d_i}^{-\alpha}}}.
	\end{align}
\end{lemma}
\begin{IEEEproof}
	See Appendix~\ref{Proof_Lemma_2}.
\end{IEEEproof}

Substituting \eqref{Eq_HandoffPr} and \eqref{Eq_Theorem_1} into \eqref{Eq_Coverage_Handoff_form2}, an upper bound on the coverage probability with handoffs can be derived, as formalized below. 
\begin{theorem}\label{Theorem_1}
	In light of the handoff probability given by \eqref{Eq_HandoffPr} and the coverage probability given by \eqref{Eq_Theorem_1}, the coverage probability with handoffs of a typical UAV is upper bounded by
	\begin{equation}\label{Theorem_Coverage_Handoff}
		\mathcal{P}_1 \leq \left[(1-\beta)+ \beta \left(1-\mathbb{P}_{\rm {H}} \right)  \right] \mathbb{P}_{\rm C_1}.
	\end{equation}
\end{theorem}

Like \cite{7967745}, if channel fading is not accounted for, it is equivalent to the case that the shape parameter of Nakagami fading goes to infinity, i.e., $m \rightarrow \infty$. In this case, an accurate expression for the coverage probability of a typical UAV can be explicitly derived. More specifically, in the absence of channel fading, the SIR given by \eqref{Eq_RxSINR} reduces to
\begin{equation}
	\eta_2 \triangleq \frac{S_2}{I_2} = \frac{\bigg|\sum\limits_{i \in \Phi_0} d^{-\frac{\alpha}{2}}_i\bigg|^2}{\sum\limits_{j \in \Phi \setminus \Phi_0} d^{-\alpha}_{j,0}}.
\end{equation}
By using the relationship between the moment generating function and the characteristic function, the characteristic function of $I_2 \triangleq \sum_{j \in \Phi \setminus \Phi_0} d^{-\alpha}_{j, 0}$ can be given by
\begin{equation}
	\Phi_{I_2}(\omega) =\exp\left(\lambda \pi {d}^2_3 - \lambda \pi {d}^2_3\, \pFq{1}{1}{-\frac{2}{\alpha}}{1-\frac{2}{\alpha}}{i\omega{d}^{-\alpha}_3} \right).
\end{equation}

By recalling \cite[Eq. (5)]{6933943}, the coverage probability of a typical UAV can be calculated as
\begin{align}\label{Eq_Coverage_NoFading-a}
	\mathbb{P}_{\rm C_2} &= \Pr\left\{\frac{S_2}{I_2} > \gamma \right\} \nonumber\\
		&= \left\{ \begin{array}{rl}
			\int\limits_{-\infty}^{\infty} \Phi_{\eta_2^{-1}}(\omega) \left(\frac{1-\exp\left(-\frac{i\omega}{\gamma}\right)}{i\omega}\right) \frac{{\rm d}\omega}{2\pi}, & \text{if } \gamma > 0; \\
			1, & \text{if } \gamma = 0,
		 	   \end{array} \right.
\end{align}
where $\Phi_{\eta_2^{-1}}(\omega)$ denotes the characteristic function of $\eta_2^{-1}$, computed by
\begin{align}
\Phi_{\eta_2^{-1}}(\omega) &= \int\limits_{H_{1}}^{H_{2}}\int\limits_{0<\bm{r} <\infty} f_{H}(x) f_{r_1,r_2,r_3}(\bm{r}) \exp\left(\lambda \pi {d}^2_3 \right. \nonumber\\
&\quad \left. - \lambda \pi {d}^2_3\, \pFq{1}{1}{-\frac{2}{\alpha}}{1-\frac{2}{\alpha}}{\frac{i\omega{d}^{-\alpha}_3}{\left(\sum\limits_{i=1}^{3}{d}^{-\frac{\alpha}{2}}_i \right)^{2}}} \right) {\rm d} \bm{r}{\rm d}x .
\end{align}
As a result, in the absence of channel fading, the coverage probability with handoffs is explicitly given by
\begin{equation} \label{Eq_Coverage_NoFading}
	\mathcal{P}_2 = \left[(1-\beta)+ \beta \left(1- \mathbb{P}_{\rm H} \right)  \right] \mathbb{P}_{\rm C_2}.
\end{equation}

\section{Dynamic Frequency Planning}
\label{Section_FrequencyAllocation}
With the higher density of BSs, ICI dominates the network performance and decreases the coverage probability computed in the previous section. Accordingly, this section develops a practical frequency planning strategy to mitigate ICI and improves network coverage probability. In particular, for mathematical tractability, the frequency allocation process is first formulated as a circle packing problem, where the entire coverage area of the network under study is filled in by multiple circles of equal size. Then, our goal is to allocate different spectrum resources to clustering triangular cells within their corresponding circle.

\subsection{Circle Packing} \label{Sec-IV-A}
On a 2D Euclidean plane, it is widely known that the highest-density lattice packing of circles is the hexagonal packing arrangement \cite[Sec. 1.4]{Conway1999}, where the centers of the circles are arranged in a hexagonal lattice, and each circle is surrounded by six other circles. In this section, we integrate the method of circle packing into our Delaunay tessellation for frequency planning. The main idea is to use a set of circles to tesselate the coverage area, each covering a cluster of adjacent triangles. The total system bandwidth is normalized to be unity and repeatedly used among different clustering triangular cells. In each cluster of triangular cells, various frequency resources are allocated to different cells. Therefore, inter-cell interference arises only in the triangular cells spanning adjacent circles. On the other hand, as the total system bandwidth is allocated on average to all the cells in a cluster, for a fixed circle area, if the circle covers fewer triangular cells, each cell will have a larger size and be allocated more spectrum resources. This frequency allocation scheme conforms to the actual spectrum demand of engineering applications.

\subsection{Frequency Reuse Distance} \label{Sec-IV-B}
To elaborate on the frequency reuse criterion, the \emph{effective interference radius}, $\varepsilon$, is defined such that each interfering BS using the same frequency band should be outside this region. The spectral efficiency is chosen as a standard performance metric for the network under study, and a predetermined minimum spectral efficiency $\mathcal{R}_{\rm th}$ in nat/sec/Hz shall be fulfilled, i.e., 
\begin{align} \label{condRate}
	\mathbb{E}\left[\ln(1 + \eta)\right]\geq \mathcal{R}_{\rm th}.
\end{align}   
Meanwhile, let $c(o, \varepsilon)$ be a circle with radius $\varepsilon$ centered at the origin $o$, $X\triangleq \left|\sum_{i \in \Phi_0}{d_i}^{- \frac{\alpha}{2}}\|{h}_i\| \right|^2$, and $Y_{\varepsilon}\triangleq \sum_{j \in {\Phi} \setminus c(o, \varepsilon)}  {d_{j, \, 0}^{-\alpha}g_j}$. It is straightforward that
\begin{align}
\mathbb{E}\left[\ln\left(1 + \frac{X}{Y_{\varepsilon}}\right)\right] 
&\leq \ln\left(1 + \mathbb{E}\left[\frac{X}{Y_{\varepsilon}}\right]\right) \nonumber\\
&\leq \ln\left(1 + 3m_1\Omega_1 \mathbb{E}\left[\frac{\sum\limits_{i=1}^3 {d_i}^{-\alpha}}{Y_{\varepsilon}}\right]\right) 
\nonumber\\
&\approx \ln\left(1 + 3m_1\Omega_1 \frac{\mathbb{E}\left[\sum\limits_{i=1}^3 {d_i}^{-\alpha}\right]}{\mathbb{E}\left[Y_{\varepsilon}\right]}\right),
\label{condRate1}
\end{align} 
which implies that a large $\varepsilon$ is desired from the received SIR viewpoint as $\mathbb{E}\left[Y_{\varepsilon}\right]$ decreases with $\varepsilon$. In contrast, as the said circle $c(o,\varepsilon)$ gets larger, there are more triangular cells in a circle, and each cell gets fewer spectrum resources. To resolve this dilemma, the optimal circular radius, $\varepsilon^{\star}$, is determined by allowing the maximum interference, that is, the inequality \eqref{condRate} takes equality: 
\begin{align} \label{condRate2}
\ln\left(1 + 3m_1\Omega_1 \frac{\mathbb{E}\left[\sum\limits_{i=1}^3 {d_i}^{-\alpha}\right]}{\mathbb{E}\left[Y_{\varepsilon^{\star}}\right]}\right)=\mathcal{R}_{\rm th},
\end{align} 
where $\mathbb{E}\left[Y_{\varepsilon^{\star}}\right]$ can be computed as \cite[Lemma~5]{6933943}
\begin{align} \label{meanInt}
	\mathbb{E}\left[Y_{\varepsilon^{\star}}\right] 
&= \int\limits_{\varepsilon^{\star}}^{\infty}  (x^2+\bar{H}^2)^{-\frac{\alpha}{2}}\mathbb{E}[g] \lambda(x) {\rm d}x \nonumber\\
&= \int\limits_{\varepsilon^{\star}}^{\infty} 2\lambda\pi m_2 \Omega_2 x (x^2+\bar{H}^2)^{-\frac{\alpha}{2}} {\rm d} x \nonumber\\
&= \frac{2}{\alpha - 2}{\lambda\pi m_2 \Omega_2 (\varepsilon^{\star 2}+{\bar{H}}^2)^{ 1-\frac{\alpha}{2}}}
\end{align} 
with $\mathbb{E}[g] = m_2 \Omega_2$ for $g\sim \Gamma(m_2, \Omega_2)$ and $\bar{H} = (H_1+H_2)/2$ referring to the average height of UAVs. Meanwhile, $\mathbb{E}[\sum_{i = 1}^3 {d_i}^{-\alpha}]$ is evaluated by
\begin{align} \label{expXevalclosedform2}
	\mathcal{M}_1(\alpha) \triangleq  \mathbb{E}\left[\sum_{i=1}^{3}d_i^{-\alpha}\right] 
		=\int_{\bm{r}>0}\left(\sum_{i=1}^{3}d_i^{-\alpha}\right)^{2}f_{r_{1},r_{2},r_{3}}(\bm{r}) \, {\rm d}\bm{r} 
\end{align}
with $\{\bm{r}: 0<r_{1}\leq r_{2}\leq r_{3} < \infty\}$.  Next, inserting \eqref{meanInt} and \eqref{expXevalclosedform2} into \eqref{condRate2}, the optimal circular radius $\varepsilon^{\star}$ for a typical UAV can be computed as
\begin{equation}\label{Eq_r_O}
\varepsilon^{\star} = \sqrt{\left(\frac{3 m_1 \Omega_1(\alpha-2)\mathcal{M}_1(\alpha)}{2\lambda\pi m_2 \Omega_2 (e^{\mathcal{R}_{{\rm th}} }-1) }\right)^{\frac{2}{2-\alpha}} -\bar{H}^2}.
\end{equation}

By definition of the circle $c(o,\varepsilon^{\star})$, a cluster of triangular cells is formed within this circle. Each cell in a given cluster uses distinct spectrum resources to eliminate intra-cluster interference. Outside the circle, $c(o,\varepsilon^{\star})$, the triangles with overlapping spectrum resources result in co-channel interference for the typical triangular cell. Therefore, the frequency reuse factor can be defined as
\begin{align} \label{freqreusefactDefinition}
\delta \triangleq \left\lfloor \frac{\mathcal{S}_{c(o,\varepsilon^{\star})}}{\mathbb{E}\left[\mathcal{S}_{\rm cell}\right]} \right\rfloor, 
\end{align}
where the numerator and denominator represent the area of the effective interfering circle and the expected area of a typical cell, respectively. In particular, the area of the circle $c(o,\varepsilon^{\star})$ is simply computed by $\mathcal{S}_{c(o,\varepsilon^{\star})} =  \pi {\varepsilon^{\star}}^2$. Furthermore, according to \cite[Table 5.11.1]{b:OkabeBoots2000} and \cite[Thm. 2.9 and Def. 2.12]{Haenggi12}, we get $\mathbb{E}\left[\mathcal{S}_{\rm cell}\right]=1/(2\lambda)$. As a result, the integer-valued frequency reuse factor in \eqref{freqreusefactDefinition} can be explicitly calculated as
\begin{align} \label{freqreusefactDefinition1}
	\delta = \left\lfloor 2\lambda\pi \varepsilon^{\star 2} \right\rfloor.
\end{align} 
Apparently, the frequency reuse factor (equivalently, the number of cells that reserve distinct spectral resources and form a cluster) grows in square law with the effective interference radius, i.e., $\delta \propto {\varepsilon^{\star}}^{2}$. 

\subsection{Frequency Allocation} \label{Sec-IV-C}
The basic idea of the proposed frequency allocation strategy is as follows. Given a finite plane of BS intensity $\lambda$, the predetermined spectral efficiency threshold is assumed to be $\mathcal{R}_{\rm th}$, and the average height of UAVs is $\bar{H}$. We first construct the Delaunay triangulation and define the corresponding triangular cells by using, e.g., the radial sweep algorithm or divide-and-conquer algorithm \cite[ch.~4]{Hjelle06}. Then, we compute the radius of circle $\varepsilon^{\star}$ as per \eqref{Eq_r_O} and then fill in the target plane with circles of radius $\varepsilon^{\star}$ by using circle packing. Given that there are $k_1$ cells in a reference circle say Circle A, the bandwidth reserved for each cell is simply $1/k_1$, where the total bandwidth for the whole system is normalized to unity. Finally, evenly assign spectrum resources for each cell according to the number of cells within each circle, starting with Circle~A. In each circle, a unique frequency band is assigned to different triangular cells. Meanwhile, the effective interference radius $\varepsilon^{\star}$ is always satisfied for each circle. In summary, the proposed frequency allocation scheme is formalized in Algorithm~\ref{alg_FrequencyAllocation}. 

\begin{algorithm} [!t]
	\caption{The Proposed Frequency Allocation Scheme}
	\small
	\label{alg_FrequencyAllocation}
	\begin{algorithmic}[1]  
		\REQUIRE The BS intensity $\lambda$, the number of antennas $M$, the path-loss exponent $\alpha$, the threshold of spectral efficiency $\mathcal{R}_{\rm th}$, the average height of UAVs $\bar{H}$.
		\STATE {Construct the Delaunay triangulation;}
		\STATE {Calculate the optimal circular radius, $\varepsilon^{\star}$, as per \eqref{Eq_r_O};}
		\STATE {Fill in the space with equal circles of radius $\varepsilon^{\star}$ by using the theory of circle packing;}
		\STATE {For each circle, count the number of triangular cells $k_i$ in the $i$-th circle, $i = 1, \cdots, n$, where $n$ denotes the total number of circles;}
		\FOR {$i := 1$ to $n$ \do} 
			\STATE {Assign spectrum resource with bandwidth $1/k_i$ to each triangular cell in the $i$-th circle;}
		\ENDFOR 
	\end{algorithmic}
\end{algorithm}

\begin{figure}[t!] 
	\centering    
	\subfloat[An illustration of frequency allocation with seven filling circles.] 
	{
		\begin{minipage}[t]{0.5\textwidth}
			\centering         
			\includegraphics[width=1.0\textwidth]{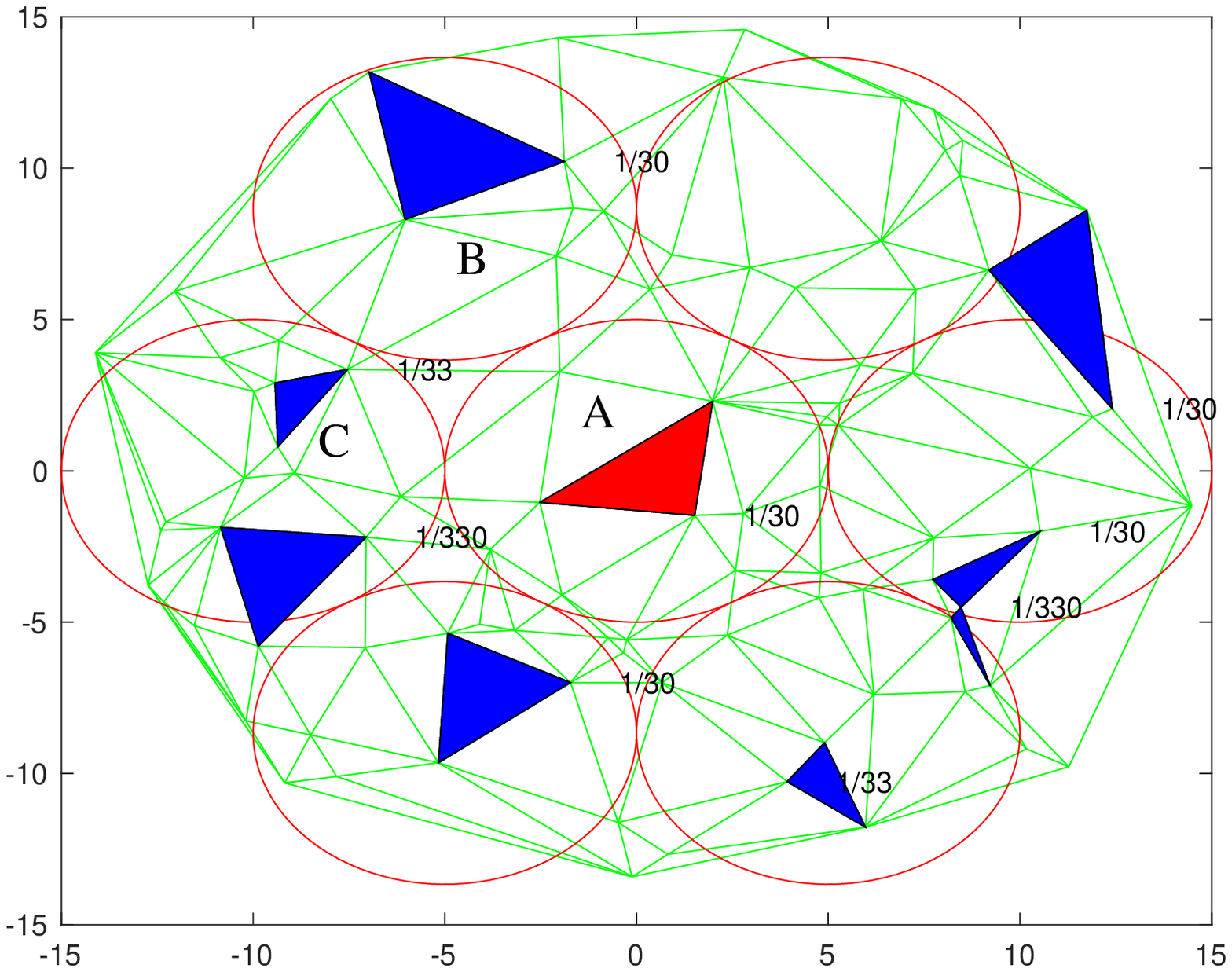}   
			\vspace{-20pt}
			\label{Fig-6a}
		\end{minipage}
	}

	\subfloat[Schematic diagram of frequency allocation.] 
	{
		\begin{minipage}[t]{0.5\textwidth}
			\centering      
			\includegraphics[width= 1.0\textwidth]{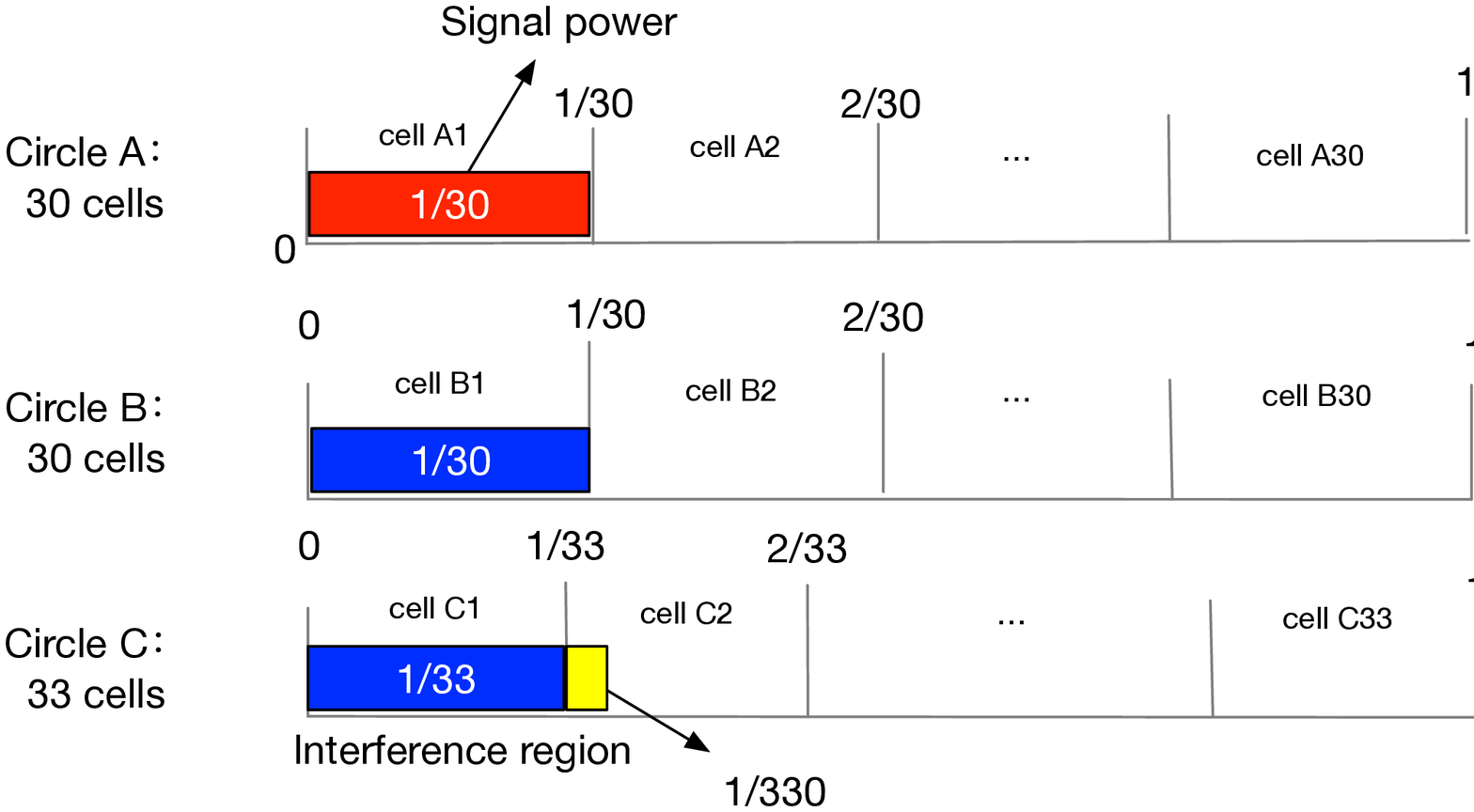}  
			\vspace{-20pt}
			\label{Fig-6b} 
		\end{minipage}
	}
	\caption{An illustrative frequency allocation scheme, where the red shaded area in Circle A denotes the typical triangular cell, and the blue shaded areas refer to the interference cells. The yellow-shaded region corresponds to the overlapping spectral region of the second interference cell in Circle C.} 
\end{figure}

For illustration purposes, Fig. \ref{Fig-6a} shows $7$ circles and $9$ shaded triangular cells sharing the same frequency band. In particular, Circle A with 30 cells is marked as the typical circle, and the red shaded area within Circle~A denotes the typical triangular cell with frequency range $[0, 1/30]$. Suppose the bandwidth of any cells in interfering circles overlaps with that of the typical cell. In that case, they are called the interfering cells, and the ratio of overlapping bandwidth is taken as a weight value to reflect the strength of interference. For instance, Circle B with 30 cells and Circle C with 33 cells stand for two interfering circles, where the blue shaded cells are the interfering ones. The corresponding overlapping bandwidth is $1/30$ for Circle B, $1/33$ for one cell and $1/30 - 1/33 = 1/330$ for another cell in Circle C. Clearly, the sum of the overlapping bandwidth of the two interfering cells in Circle C equals the bandwidth of the typical red cell in Circle~A. 

For a more intuitive understanding, Fig.~\ref{Fig-6b} gives a schematic diagram of frequency allocation. As said before, assuming the first cell A1 in Circle A to be the typical cell, the cell B1 in Circle B and the cell C1 and part of the cell C2 in Circle C have an overlapping spectrum with the typical cell A1. Therefore, cells B1, C1, and C2 introduce ICI to cell A1. To distinguish the proportions of the overlapping spectrum of different cells in the same Circle, blue and yellow shaded regions are used to mark the corresponding overlapping spectrum regions in Circle C.

On the other hand, Fig.~\ref{Fig-6a} also discloses that there are gaps between tangential filling circles in the proposed circle-filling method. To evaluate the efficiency of the circle packing strategy, we are concerned with the smallest number of circles if seamless coverage is necessary. Suppose that we are given a subset $\mathcal{K} \subset \mathbb{R}^2$ and asked to seamlessly cover it by circles of a given radius $\varepsilon > 0$. Assume the subset $\mathcal{K}$ can be seamlessly covered by $N_C$ circles with radii $\varepsilon$. Comparing their areas, we have $|\mathcal{K}| \leq N_C | \mathcal{C}(0, \varepsilon)|$, where $| \cdot |$ denotes the area in $\mathbb{R}^2$ and $\mathcal{C}(0, \varepsilon)$ is a Euclidean circle centered at the origin with radius $\varepsilon$. Dividing both sides by $|\mathcal{C}(0, \varepsilon)|$ yields the lower bound $N_C \geq |\mathcal{K}|/|\mathcal{C}(0, \varepsilon)|$. On the other hand, we can construct $N_P$ closed disjoint circles $\mathcal{C}(x_i, \varepsilon/2)$ with centers $x_i \in \mathcal{K}$ and radii $\varepsilon/2$. While these circles may not need to fit entirely into $\mathcal{K}$, they do fit into a slightly inflated set, namely $\mathcal{K} + \mathcal{C}(0, \varepsilon/2)$. Comparing their areas, we obtain $N_P |\mathcal{C}(0, \varepsilon/2)| \leq |\mathcal{K} + \mathcal{C}(0, \varepsilon/2)|$, which leads to the upper bound $N_P \leq |\mathcal{K} + \mathcal{C}(0, \varepsilon/2)|/|\mathcal{C}(0, \varepsilon/2)|$. Finally, by recalling \cite[Lemma 4.2.8]{2018High}, we have $N_C \leq N_P$, yielding the following theorem.
\begin{theorem}[Bounds on the number of filling circles]
	Let $\mathcal{K}$ be a subset of $\mathbb{R}^2$ and $\varepsilon > 0$, assuming $\mathcal{K}$ can be seamlessly covered by $N_C$ circles with radius $\varepsilon$, then
	\begin{equation} \label{Eq-covering-a}
		\frac{|\mathcal{K}|}{|\mathcal{C}(0, \varepsilon)|} \leq N_C \leq \frac{|\mathcal{K}+ \mathcal{C}(0, \frac{\varepsilon}{2})|}{|\mathcal{C}(0, \frac{\varepsilon}{2})|}.
	\end{equation}
	If $\mathcal{K}$ is a unit circle $\mathcal{C}(0, 1)$, then \eqref{Eq-covering-a} reduces to
	\begin{equation} \label{Eq_N_C_Unit}
		\left(\frac{1}{\varepsilon} \right)^2 \leq N_C \leq \left(\frac{2}{\varepsilon} +1\right)^2.
	\end{equation}
\end{theorem}

{Suppose that we fill the unit circle $\mathcal{C}(0, 1)$ with different radii ($\varepsilon$) of filling circles. The upper and lower bounds on the number of filling circles for seamless coverage, computed by \eqref{Eq_N_C_Unit}, are listed in Table \ref{tab1}, compared with the simulation results of the number of filling circles for the seamless coverage and the number of filling circles for the highest-density lattice packing described in Section~\ref{Sec-IV-A}. As expected, the simulation results pertaining to the seamless coverage range between the upper and lower bounds. However, when the radius of filling circles is small (e.g., $\varepsilon = 0.025$ or $\varepsilon = 0.05$ in Table~\ref{tab1}), the number of filling circles for the lattice packing is even smaller than the lower bound computed by \eqref{Eq_N_C_Unit}. This observation is of no surprise because the highest-density lattice packing allows a bit of gap among filling circles (rather than seamless coverage), as illustrated in Fig.~\ref{Fig-6a}. When the radius of filling circles becomes larger, the gap becomes smaller such that the numbers of filling circles for the seamless coverage and the lattice packing coincide, as shown in Table~\ref{tab1}.}

\begin{table}[h]
	\centering
	\caption{The number of filling circles}	
	\label{tab1}
	\setstretch{1.25}	
	\scalebox{0.7}{
	\begin{tabular}{!{\vrule width1.2pt} c !{\vrule width1.2pt} c|c|c|c|c|c|c|c !{\vrule width1.2pt}}
		\Xhline{1.2pt}  
		Radii of filling circles ($\varepsilon$)  & 0.025 & 0.05 & 0.1&0.2&0.3&0.4&0.5&0.6\\
		\hline   
		Upper bounds& 6561 & 1681 & 441&121&58.8&36&25&18.8\\
		\hline     	
		Seamless coverage &  2029& 517 &  151 &  37&  19 & 13 & 7&  7\\
		\hline       		                                   
		{Lattice packing} & 1519& 397 & 109 &32& 19 &13 &7&7\\
		\hline   
		Lower bounds& 1600 & 400 & 100&25&11.1&6.3&4&2.8\\
		\Xhline{1.2pt}
	\end{tabular}
}
\end{table}

\subsection{Coverage Probability After Frequency Allocation}  \label{Sec-IV-D}
After performing the frequency allocation mentioned above, the interfering BSs, not all BSs in $\Phi$, which transmit in the same frequency band, are a thinned version of the original PPP with intensity $\lambda' = \lambda/\delta$. Since a thinned version of a PPP is again a PPP, the coverage probability with handoffs can be calculated as per \eqref{Theorem_Coverage_Handoff} for a typical UAV with $\lambda^{\prime}$ in place of~$\lambda$.

\section{Simulation Results and Discussions}
\label{Section_Simulation}
This section illustrates numerical results computed per the previously obtained analytical expressions are illustrated and compared with extensive Monte-Carlo simulation results.  In the pertaining simulation experiments, a large-scale wireless network with BS density $20 \ {\rm BSs}/{\rm km}^2$ is assumed. The BS height is set to $H_{\rm BS} = 25$ m and the Ricean fading factor is set to $K = 1$. 

\begin{figure}[!t]
	\centering
	\includegraphics[width=3.5in, clip]{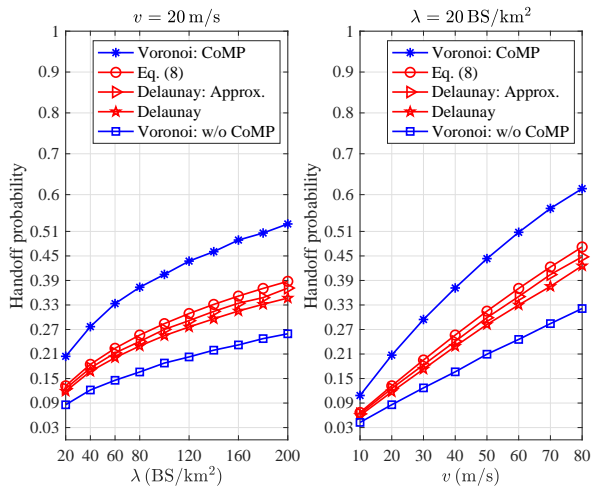}
	\caption{Handoff probabilities versus the BS density (left) as well as the average moving speed of UAVs (right), under the proposed Poisson-Delaunay triangulation, the Poisson-Voronoi tessellation with three nearest cooperating BSs and the same scheme without CoMP. The maximum height of UAV is set to $H_2 = 70 \text{ m}$.}
	\label{Fig-7}
\end{figure}

\subsection{Handoff Probability}
Figure~\ref{Fig-7} depicts the handoff probabilities of UAVs under the proposed Poisson-Delaunay triangulation and the traditional Poisson-Voronoi tessellation with three nearest cooperating BSs, which are illustrated with legends ``Delaunay" and ``Voronoi: CoMP", respectively. For comparison purposes, Poisson-Voronoi tessellation without CoMP is also considered, depicted with legend ``Voronoi: w/o CoMP". Finally, the results with legend ``Delaunay: Approx." correspond to the simulation results of handoff probability using the equivalent Voronoi approximation method described in Subsection~\ref{Subsection-Handoff}. Specifically, the left panel of Fig.~\ref{Fig-7} depicts the handoff probability versus the BS density~$\lambda$. It is observed that, for a fixed moving speed of UAVs ($v = 20 \ {\rm m/s}$), the handoff probability monotonically increases with $\lambda$. Similarly, the right panel of Fig.~\ref{Fig-7} shows the handoff probability monotonically increases with the moving speed of UAVs for a fixed BS intensity. These observations are because handoffs occur more frequently as the cell size becomes smaller or UAVs fly faster, as expected. On the other hand, the handoff probability of our proposed model is much lower than that of the Voronoi model with three nearest cooperating BSs, especially for large $\lambda$ or high $v$. For instance, it is seen from the right panel of Fig.~\ref{Fig-7} that, given $\lambda = 20\, {\rm BS/km^2}$ and $v = 40 \, {\rm m/s}$, the handoff probabilities of the traditional CoMP scheme and the proposed scheme are about $37\%$ and $24\%$, respectively, thus illustrating the superiority of the proposed scheme. The fundamental reason for the superiority is that, in our scheme, a handoff occurs only if the UAV's average received power changes, while the distances between a UAV and its three serving BSs are not necessarily the closest. However, in the Poisson-Voronoi network with three nearest BSs, a handoff occurs if any of the three nearest BSs changes, thus resulting in a higher handoff probability. 

Figure~\ref{Fig-7} also shows that the handoff probability of the proposed CoMP scheme is a bit higher than that of the Voronoi scheme without CoMP. This phenomenon is not surprising as the average area of Delaunay triangular cells is half of their dual Voronoi cells \cite{Xia2018Un}. On the other hand, it is observed that the simulation results based on the equivalent Voronoi approximation method match well with the simulation results, demonstrating this method's feasibility.  Finally, it is seen that the numerical results computed as per \eqref{Eq_HandoffPr} agree well with the simulation ones, which corroborates the effectiveness of our analysis. 

\begin{figure}[!t]
	\centering
	\includegraphics[width=3.5in, clip]{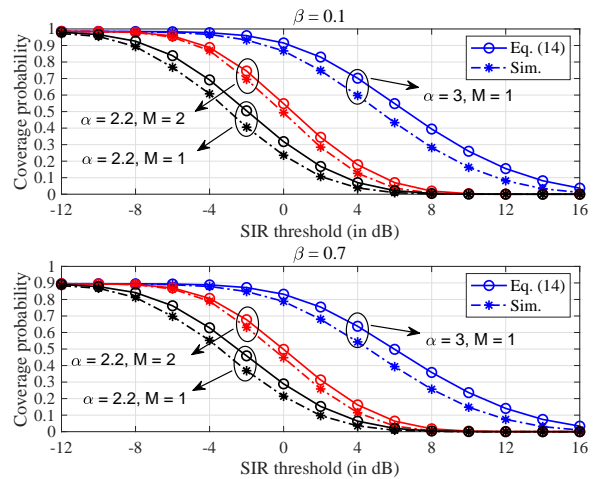}
	\caption{Coverage probability with handoffs versus SIR threshold in the unit of dB with $\alpha = 2.2$ or $3$ for a typical UAV, with an average moving speed of $20$ m/s. }
	\label{Fig-8}
\end{figure}

\subsection{Coverage Probability with Handoffs}
In this subsection, simulation results about the coverage probability with handoffs are firstly depicted, then exact numerical results about the coverage probability in the absence of fading are shown. Next, the effect of varying heights of UAVs is discussed, and the coverage probability after frequency allocation is finally illustrated.

\subsubsection{Coverage Probability with Handoffs}
Figure~\ref{Fig-8} shows the coverage probability with handoffs versus the SIR threshold $\gamma$ in dB for a typical UAV, where the top panel corresponds to the case of $\beta = 0.1$ while the bottom panel to the case of $\beta = 0.7$, where $\beta$ reflects the probability of connection failure due to handoffs, as defined immediately after Eq.~\eqref{Eq_Coverage_Handoff_form2}. It can be observed that the coverage probability with $\beta = 0.1$ is higher than that with $\beta = 0.7$ as the system is less sensitive to handoff with smaller $\beta$. On the other hand, for either case, the coverage probability decreases with a higher SIR threshold while increasing with both more prominent path-loss exponent $\alpha$ and more antennas $M$, as expected.

\subsubsection{Special Case in the Absence of Fading}
\label{Sec_SImulated_Coverage}
\begin{figure}[!t]
	\centering
	\includegraphics[width=3.5in, clip]{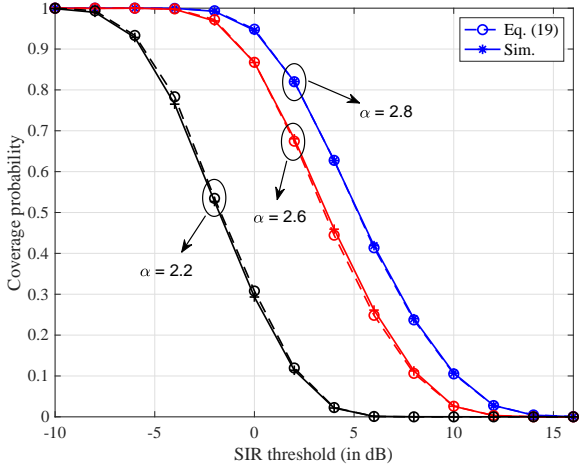}
	\caption{Coverage probability with handoffs in the absence of fading under different values of path-loss exponent $\alpha$, with $\beta = 0.1$, ${v} = 9$ m/s, $H_1 = 30$ m, and $H_2 = 70$ m.}
	\label{Fig-9}
\end{figure}

Figure~\ref{Fig-9} illustrates the coverage probability with handoffs versus the SIR threshold in the absence of fading, with different path-loss exponent $\alpha$. It is seen that the numerical results computed as per \eqref{Eq_Coverage_NoFading} match well with the corresponding simulated ones, which verifies the accuracy of our analysis. Besides, the coverage probability increases with the path-loss exponent $\alpha$. The reason behind this observation is that, while raising $\alpha$ decreases the desired signal, it more significantly decreases the interference power by recalling the fact that the interfering signals are usually farther from the BS than the desired signal, thus increasing the whole SIR and the coverage probability. 

\begin{figure}[!t]
	\centering
	\includegraphics[width=3.5in, clip]{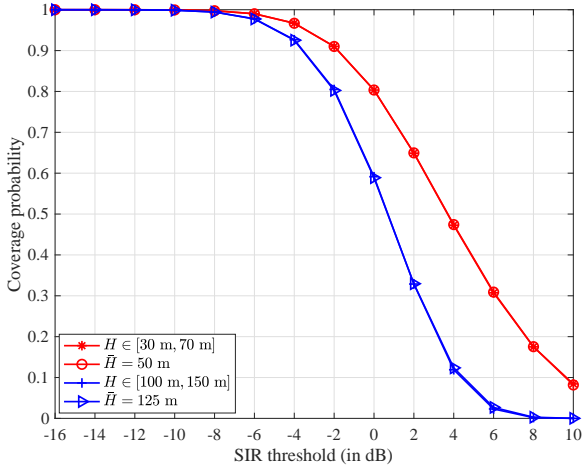}
	\caption{Coverage probability of a typical UAV with varying heights, with $\beta = 0$, $\alpha = 2.6$ and $\lambda = 20 \ \mathrm{BS/km}^2$.}
	\label{Fig-10}
\end{figure}

\subsubsection{The Average Height of UAVs}
According to the principle of circular packing discussed in Subsection~\ref{Sec-IV-A}, the radius of circles depends on the instantaneous height (i.e., $H$) of a typical UAV. For ease of frequency planning, the optimal circular radius is calculated concerning its average height (i.e., $\bar{H}$), as given by Eq.~\eqref{Eq_r_O}. Figure~\ref{Fig-10} plots the coverage probabilities computed by the instantaneous heights and average ones. It is seen that they match well with each other for either the case of $H \in [30~\mathrm{m}, 70~\mathrm{m}]$ with $\bar{H} = 50~\mathrm{m}$ or the case of $H \in [100~\mathrm{m}, 150~\mathrm{m}]$ with $\bar{H} = 125~\mathrm{m}$. The same observation can also be made from \cite[Fig. 9]{7967745}. As the radius of the terrestrial network coverage is much larger than the height of UAVs, the small variation of UAV height does not significantly impact the coverage probability. Also, it is observed that the coverage probability decreases with a higher average height of UAVs, as expected. Therefore, the average height of UAVs can serve as a key parameter for effective frequency planning and other performance evaluation.

\subsubsection{Coverage Probability After Frequency Allocation}
\begin{figure}[!t]
	\centering
	\includegraphics[width=3.5in, clip]{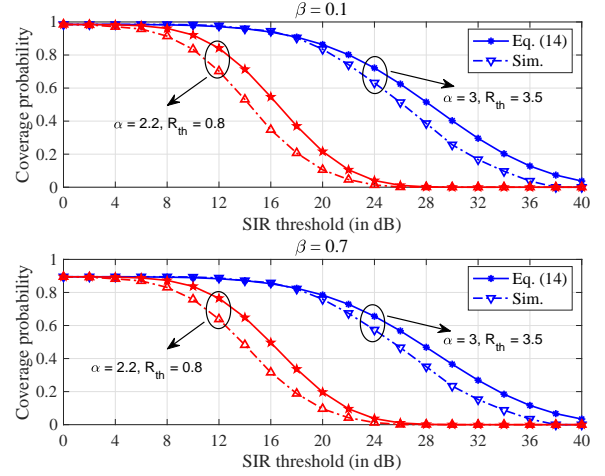}
	\caption{Coverage probability versus SIR threshold in the unit of dB under different  $\mathcal{R}_{{\rm th}}$ and $\alpha$, with $v = 20$ m/s.}
	\label{Fig-11}
\end{figure}

Figure~\ref{Fig-11} depicts the coverage probability with handoffs versus the SIR threshold $\gamma$, under the path-loss exponent $\alpha = 3$ for a given $\mathcal{R}_{{\rm th}} = \ln(1+10^{1.5}) = 3.5$ nat/sec/Hz or $\alpha = 2.2$ for $\mathcal{R}_{{\rm th}} = 0.8$ nat/sec/Hz. The corresponding frequency
reuse factor is calculated via \eqref{freqreusefactDefinition1}. Compared with Fig.~\ref{Fig-8}, it can be seen that, for a fixed SIR threshold, the coverage probability increases significantly. This is because the interference is reduced effectively after dynamic frequency allocation. 

\subsection{Comparison with Voronoi Tessellation}
\begin{figure}[t!] 
	\centering    
	\subfloat[Coverage probability without handoffs.] 
	{
		\begin{minipage}[t]{0.5\textwidth}
			\centering         
			\includegraphics[width=1.0\textwidth]{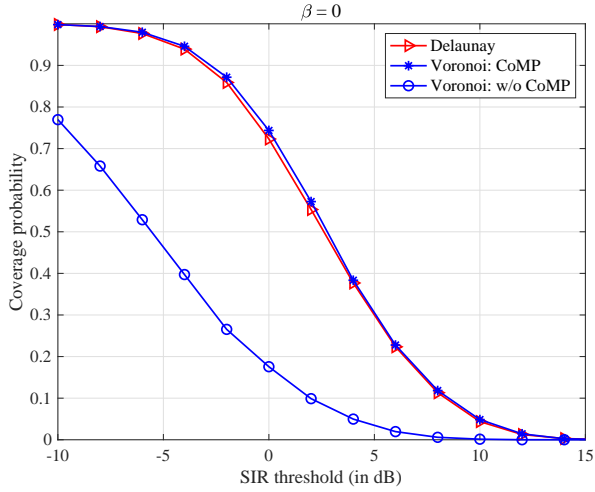}   
			\vspace{-10pt}
			\label{Fig_12a}
		\end{minipage}
	}

	\subfloat[Coverage probability with handoffs] 
	{
		\begin{minipage}[t]{0.5\textwidth}
			\centering      
			\includegraphics[width= 1.0\textwidth]{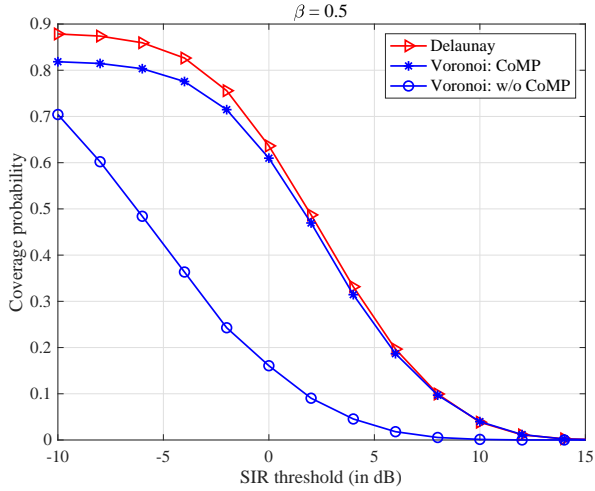}  
			\vspace{-10pt}
			\label{Fig_12b} 
		\end{minipage}
	}
	\caption{Coverage probability versus SIR threshold in the unit of dB, under the proposed Delaunay CoMP scheme and the Poisson-Voronoi tessellation scheme without CoMP and the case with three nearest cooperating BSs on the left, with $\alpha = 2.6$ and ${v}=40$ m/s.} 
	\label{Fig-12}  
\end{figure}

Figure~\ref{Fig-12} illustrates the coverage probability of a typical UAV versus the SIR threshold $\gamma$ in dB, regarding the proposed Delaunay CoMP scheme and the Poisson-Voronoi tessellation scheme with three nearest cooperating BSs. For comparison purposes, the simulation results of the Poisson-Voronoi tessellation without CoMP are also plotted. If no handoff is accounted for (i.e., $\beta = 0$), Fig.~\ref{Fig_12a} shows that the proposed Delaunay scheme achieves a slightly lower coverage probability than the Voronoi scheme with three cooperating BSs. This is because the latter always chooses the three nearest serving BSs through exhaustive searching. However, when handoff is accounted for, for instance, in the case of $\beta = 0.5$, Fig.~\ref{Fig_12b} shows that the coverage probability of our proposed Delaunay scheme outperforms that of the Voronoi scheme with three nearest cooperating BSs, especially at low SIR threshold. This comparison concludes that our CoMP scheme is preferable as UAVs inevitably require handoffs. 

\section{Concluding Remarks}
\label{Section_Conclusion}
This paper developed a coordinated multi-point (CoMP) transmission scheme for ground-to-air communications beyond 5G using the principle of stochastic geometry regarding Poisson-Delaunay triangulation. In particular, each UAV is jointly served by three terrestrial BSs, thus enhancing the communication quality. Also, a dynamic frequency allocation algorithm was designed to eliminate inter-cell interference using circle packing. After deriving the handoff probability and coverage probability for a typical UAV, simulation and numerical results demonstrated that the proposed CoMP transmission outperforms the conventional Voronoi scheme in terms of handoff probability and coverage probability. Consequently, the proposed CoMP scheme is promising for ground-to-air communications where frequent UAV handoffs occur.

\appendices
\section{Proof of Lemma~\ref{Lemma_1}}
\label{Proof_Lemma_1}
\begin{figure}[t!] 
	\centering    
	\subfloat[${v}\cos(\phi_k) \leq r$.] 
{
		\begin{minipage}[t]{0.25\textwidth}
			\centering         
			\includegraphics[width=0.9\textwidth]{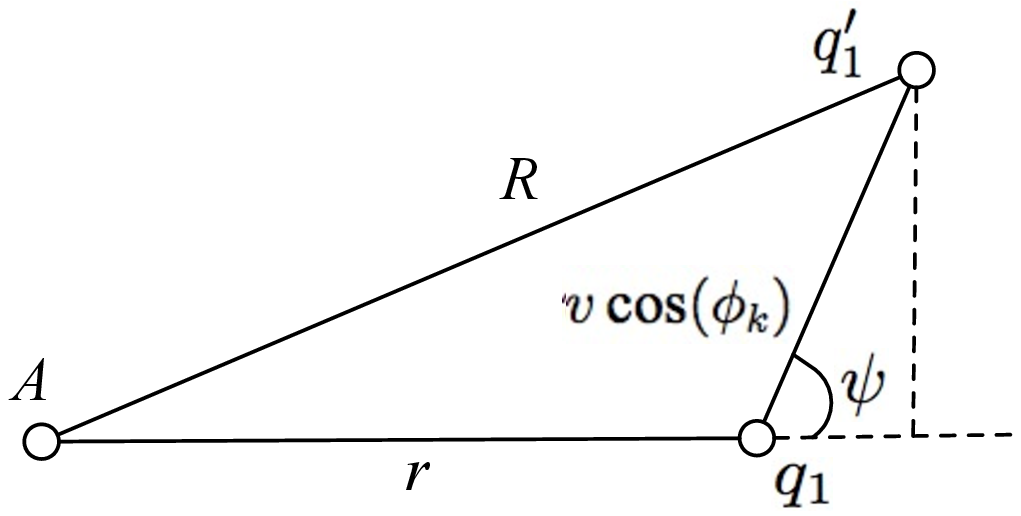}   
			\label{Fig-13a}
		\end{minipage}
} 
	\subfloat[${v}\cos(\phi_k) > r$.] 
{
		\begin{minipage}[t]{0.25\textwidth}
			\centering      
			\includegraphics[width= 0.8\textwidth]{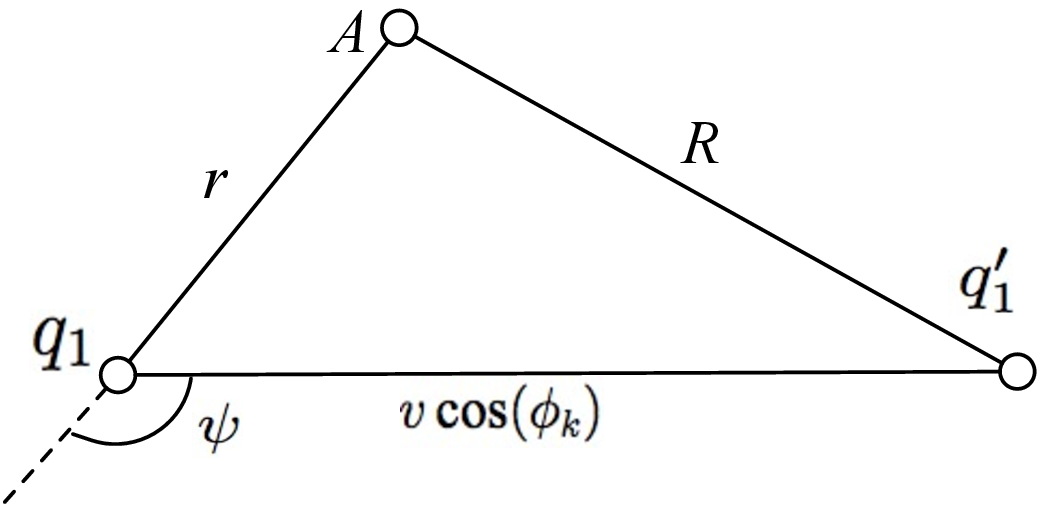}  
			\label{Fig-13b} 
		\end{minipage}
}
	\caption{Relationship between $r$, ${v}\cos(\phi_k)$, and $R$.} 
	\label{Fig-13}  
\end{figure}

Given $\{\rho_{k},H_{k-1}, H_k\}$, the conditional handoff probability can be computed as
\begin{align}
	\mathbb{P}_{\rm H|\{\rho_{k},H_{k-1}, H_k \}}
		&= \Pr\left\{   \Phi(c(q_{1}^{\prime},R)\setminus c(q_{1},r) >0|\rho_{k},H_{k-1}, H_k  )  \right\} \nonumber\\ &=1-\exp(-2\lambda | c(q_{1}^{\prime},R)\setminus c(q_{1},r)|),
\label{Eq_CondotionalHandoffPr_1}
\end{align}
where \eqref{Eq_CondotionalHandoffPr_1} comes from the void probability of PPP, and $c(q_{1}^{\prime}, R)$ denotes a circle with radius $R$ centered at $q_{1}^{\prime}$. Also, the coverage area when the UAV moves from $q_{1}$ to $q_1^{\prime}$ can be computed by:
\begin{equation}\label{Eq_CommomArea}
	|c(q_{1}^{\prime},R)\setminus c(q_{1},r)| = |c(q_{1}^{\prime},R)| -|c(q_{1}^{\prime},R) \cap   c(q_{1},r)|.
\end{equation}
Due to the symmetry, $\psi$ is considered to be uniformly distributed in the range $[0,\pi]$. By using the relationship between $r$ and $v\cos(\phi_k)$, there are two different scenarios as shown Fig.~\ref{Fig-13}. In particular, Fig.~\ref{Fig-13a} refers to the case of $v\cos(\phi_k) \leq r$ where $\psi \in [0, {\pi}/{2}]$ and Fig.~\ref{Fig-13b} represents the case of $v\cos(\phi_k) > r$ where $\psi \in [{\pi}/{2}, \pi]$.

As shown in Fig.~\ref{Fig-13a}, if $v\cos(\phi_k) \leq r$, the common area between two intersecting circles with radius $r$ and $R$, respectively, can be calculated as: 
\begin{align}
	\lefteqn{|c(q_{1}^{\prime},R) \cap   c(q_{1},r)|} \nonumber\\
	&= r^2 \angle q_{1}^{\prime}q_{1}A + R^2 \angle A q_1^{\prime} q_{1} - {v}\cos(\phi_t)r\sin(\angle q_{1}^{\prime}q_{1}A).
\end{align}
From Fig.~\ref{Fig-13a}, we have
$\angle q_{1}^{\prime}q_{1}A = \pi - \psi$ and $\angle A q_1^{\prime} q_{1} = \psi - \arcsin\left({{v}\cos(\phi_k)\sin(\psi)}/{R}  \right)$, which implies
\begin{align}\label{Eq_CommomArea_2}
|c(q_{1}^{\prime},R) \cap   c(q_{1},r)| 
&=  R^2(\psi - \arcsin\left({{v}\cos(\phi_k)\sin(\psi)}/{R}  \right)  \nonumber\\
&\quad + r^2(\pi - \psi) - {v}\cos(\phi_k)r\sin(\psi).
\end{align}
Substituting \eqref{Eq_CommomArea} and \eqref{Eq_CommomArea_2} into \eqref{Eq_CondotionalHandoffPr_1} yields the first line of \eqref{Eq_CondotionalHandoffPr_2}.

As shown in Fig.~\ref{Fig-13b}, if $v\cos(\phi_k) > r$, i.e., when $\psi$ is between $0$ and ${\pi}/{2}$ or ${v}\cos(\phi_k) \cos(\pi -\psi) \leq r$, $\angle q_{1}Aq_1^{\prime} = \arcsin\left({{v}\cos(\phi_k)\sin(\psi)}/{R}  \right)$ is between $0$ and ${\pi}/{2}$, \eqref{Eq_CommomArea_2} still holds. However, when $\angle q_{1}Aq_1^{\prime}$ is between ${\pi}/{2}$ and $\pi$,  $\psi$ is between ${\pi}/{2}$ and $\pi$ and ${v}\cos(\phi_k) \cos(\pi -\psi) > r$, $\arcsin\left({{v}\cos(\phi_k)\sin(\psi)}/{R}  \right)$ in \eqref{Eq_CommomArea_2} needs replaced by $\pi -\arcsin\left({{v}\cos(\phi_k)\sin(\psi)}/{R} \right)$, yielding the second line of \eqref{Eq_CondotionalHandoffPr_2}. This completes the proof.

\section{Proof of Lemma~\ref{Lemma_2} }
\label{Proof_Lemma_2}
By using Cauchy-Schwarz’s inequality, an upper bound on the coverage probability given by \eqref{Eq_Theorem_1} can be explicitly calculated as
\begin{align}
\mathbb{P}_{\rm C_1} \hspace{-3pt} &= \mathbb{E}\left[ \Pr \left\{ \eta > \gamma\right\} \right] \nonumber \\
	&\leq \int\limits_{H_1}^{H_2}\int\limits_{0 < \bm{r}< \infty}  \Pr \left\{\frac{\left(d_1^{-\alpha} + d_2^{-\alpha} + d_3^{-\alpha}\right) \sum_{i=1}^{3}g_i }{I} > \gamma |d_i\right\} \nonumber\\
	&\quad \times  f_{H}(x) f_{r_1, r_2, r_3}(\bm{r}) \, {\rm d}\bm{r} \, {\rm d}x\label{Eq_Coverage_O_a}\\
	& = \int\limits_{H_1}^{H_2} \int\limits_{0 < \bm{r} < \infty} \Pr \left\{\sum_{i=1}^{3}g_i > \frac{\gamma I}{d_1^{-\alpha} + d_2^{-\alpha}+ d_3^{-\alpha}}   \right\} \nonumber\\
	&\quad \times  f_{H}(x) f_{r_1, r_2, r_3}(\bm{r}) \, {\rm d}\bm{r} \, {\rm d}x  \nonumber\\
	&=\hspace{-5pt}\int\limits_{H_1}^{H_2} \int\limits_{0 <\bm{r}< \infty}\hspace{-10pt} f_{H}(x) f_{r_1, r_2, r_3}(\bm{r}) \sum_{k=0}^{\varpi-1}\frac{1}{k!}\left(\frac{\gamma}{\Omega_1}\left(\sum\limits_{i=1}^{3}d_i^{-\alpha}\right)^{-1}  \right)^k \nonumber\\
	&\quad \times \mathbb{E}_{I}\left[I^k \exp\left(-\frac{\gamma I}{\Omega_1}\left(\sum\limits_{i=1}^{3}d_i^{-\alpha}\right)^{-1} \right)\right] \, {\rm d}\bm{r} \, {\rm d}x,
\label{Eq_Coverage_O_b}
\end{align} 
where \eqref{Eq_Coverage_O_b} follows from the expansion of the complementary cumulative probability density function (CCDF) of $\sum_{i=1}^{3}g_i$, which is obviously the sum of three Gamma random variables, given by
\begin{align}
F_{\sum_{i=1}^{3}g_i}(x) 
	&= 1- \frac{1}{\Gamma(3m_1)}\gamma\left(3m_1, \frac{x}{\Omega_1}\right) \nonumber\\
	&= \sum\limits_{k=0}^{\varpi-1}\frac{1}{k!}\left(\frac{x}{\Omega_1}\right)^k \exp\left(-\frac{ x}{\Omega_1}\right), \label{CCDF_U1_b}
\end{align}
with $\varpi \triangleq {\rm round} (3m_1)$. Then, by recalling the relationship between the moments and Laplace transform of a random variable, \eqref{Eq_Coverage_O_b} can be derived as 
\begin{align}\label{Eq_Theorem_Derivative}
\mathbb{P}_{\rm C_1}
&\leq \int\limits_{H_1}^{H_2}\int\limits_{0<\bm{r} <\infty} \hspace{-10pt} f_{H}(x) f_{r_{1}, r_{2}, r_{3}}(\bm{r}) \sum_{k=0}^{\varpi-1}\frac{1}{k!}\left(- \frac{ \gamma }{\Omega_1 \left(\sum\limits_{i=1}^{3}d_{i}^{-\alpha}\right)}\right)^k \nonumber\\
&\quad \times  \frac{\partial^k L_{I_{1}}(s)}{\partial s^k} \bigg|_{s=\frac{ \gamma }{\Omega_1 \left(\sum\limits_{i=1}^{3}d_{i}^{-\alpha}\right)}} \, {\rm d}\bm{r} \, {\rm d}x,
\end{align}
where $L_{I_1}(s)$ can be explicitly computed as 
\begin{align}
&L_{I_1}(s) \nonumber\\
&= \mathbb{E}_{{\Phi \setminus \Phi_0}}\left[ \prod_{j \in \hat{\Phi}} \mathbb{E}_{g_j}\left[ \exp\left(- s d_{j, \, 0}^{-\alpha} g_j  \right)  \right]   \right]   \\
&= \exp \left( \hspace{-3pt} -2\lambda \pi \int\limits_{r_3}^{\infty} \hspace{-5pt}\left( 1- \mathbb{E}_{g} \hspace{-3pt} \left[\exp  \hspace{-3pt} \left(-s g (\sqrt{x^2+H^2})^{-\alpha} \right) \right]x{\rm d}x  \right)\right)\nonumber\\
&= \exp\left(\lambda \pi {d_3}^2+\frac{2}{\alpha}\lambda \pi s^{\frac{2}{\alpha}}\mathbb{E}_{g}\left[h^{\frac{2}{\alpha}}_1 \gamma\left(-\frac{2}{\alpha}, s {d_3}^{-\alpha}g\right)\right]\right)  \label{Eq_Eta_a}\\
&= \exp\left(\lambda \pi {d_3}^2 - \lambda \pi {d_3}^2 \mathbb{E}_{g}\left[\pFq{1}{1}{-\frac{2}{\alpha}}{1-\frac{2}{\alpha}}{-s{d_3}^{-\alpha}g}\right]\right) \label{Eq_Eta_b}\\
&= \exp\left(\lambda \pi {d_3}^2 - \lambda \pi {d_3}^2 \pFq{2}{1}{m_2,-\frac{2}{\alpha}}{1-\frac{2}{\alpha}}{-s {d_3}^{-\alpha} {\Omega_2}}\right), \label{Eq_Eta_c}
\end{align} 
in which \cite[8.351]{Gradshteyn00} is exploited to reach \eqref{Eq_Eta_b}, and \eqref{Eq_Eta_c} follows the fact $g_j \sim \Gamma(m_2,\Omega_2)$.
 
Finally, by using a similar method to that in \cite[Appendix~A]{8976426}, the recursive relations between the derivatives of $L_{I_1}(s)$ can be attained, yielding \eqref{Eq_Theorem_1}.

\bibliographystyle{IEEEtran}
\bibliography{References}

\vfill
\begin{IEEEbiography}
	[{\includegraphics[width=1in, height=1.25in, clip, keepaspectratio]{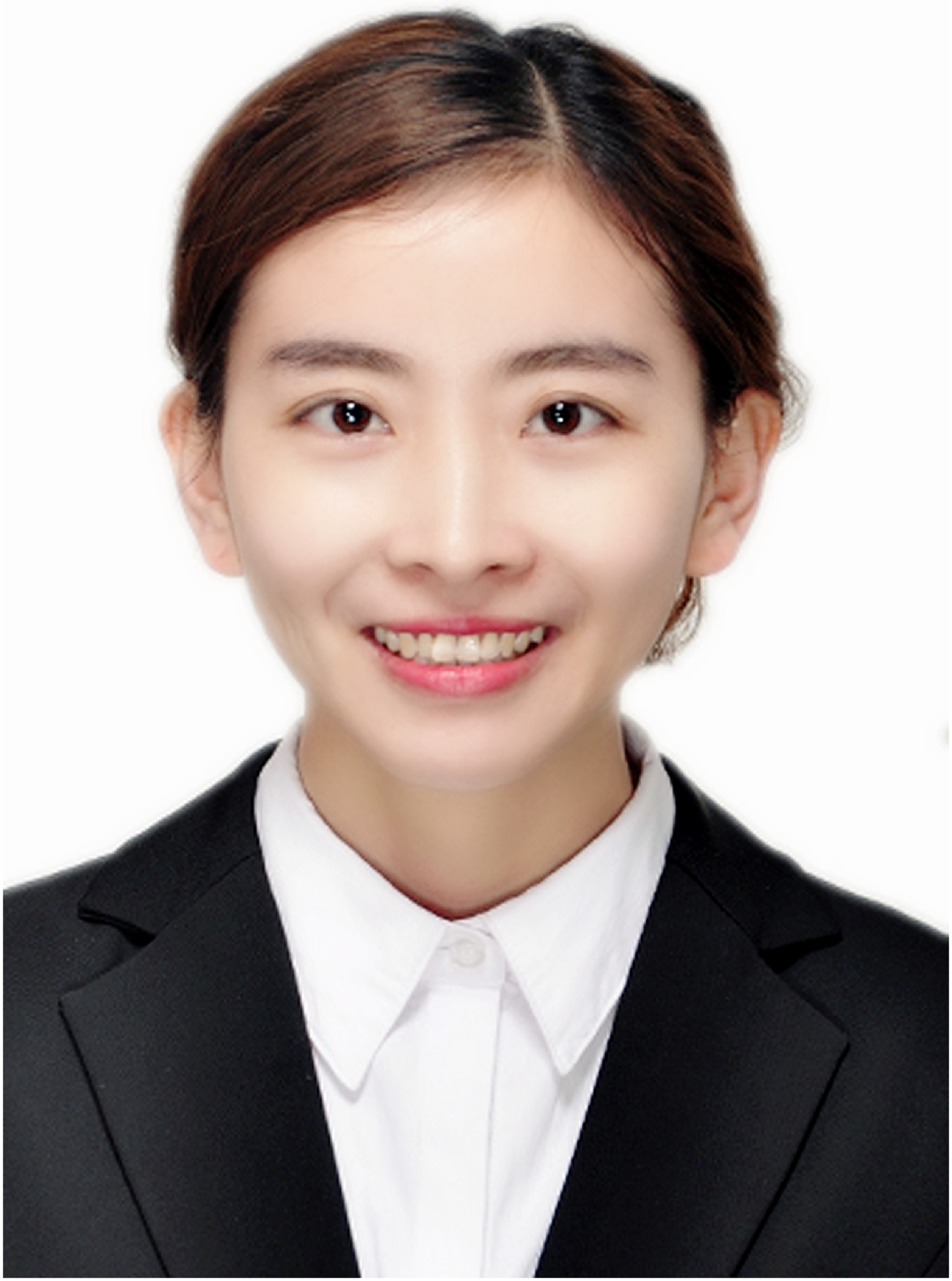}}]{Yan Li}  received a B.S. degree in Electronic Information Engineering from Hunan Normal University, Changsha, China, in 2013, and the M.S. degree in Electronics and Communication Engineering from Sun Yat-sen University, Guangzhou, China, in 2016. She received her PH.D. degree in Information and Communication Engineering at Sun Yat-sen University, Guangzhou, China, in 2021. She is currently a Lecturer at the School of Computer and Communication Engineering, Changsha University of Science and Technology, Changsha, China. Her research interests include modeling and analysis of cellular networks and cooperative communications based on stochastic geometry theory.
\end{IEEEbiography}

\begin{IEEEbiography}
	[{\includegraphics[width=1in, height=1.25in, clip, keepaspectratio]{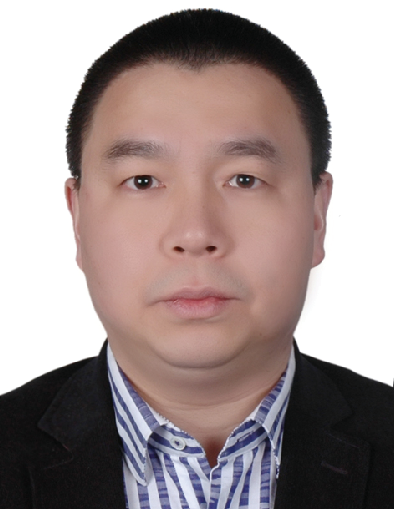}}]{Minghua Xia} (Senior Member, IEEE) received the Ph.D. degree in Telecommunications and Information Systems from Sun Yat-sen University, Guangzhou, China, in 2007. 
	
	From 2007 to 2009, he was with the Electronics and Telecommunications Research Institute (ETRI) of South Korea, Beijing R\&D Center, Beijing, China, where he worked as a member and then as a senior member of the engineering staff. From 2010 to 2014, he was in sequence with The University of Hong Kong, Hong Kong, China; King Abdullah University of Science and Technology, Jeddah, Saudi Arabia; and the Institut National de la Recherche Scientifique (INRS), University of Quebec, Montreal, Canada, as a Postdoctoral Fellow. Since 2015, he has been a Professor at Sun Yat-sen University. Since 2019, he has also been an Adjunct Professor with the Southern Marine Science and Engineering Guangdong Laboratory (Zhuhai). His research interests are in the general areas of wireless communications and signal processing. 
\end{IEEEbiography}
\end{document}